\documentclass[twocolumn]{aastex631}
\usepackage{amsmath}

\shorttitle{CEE outcome II.}
\shortauthors{Ge et al.}
\graphicspath{{./}}

\begin{document}

\title{The Common Envelope Evolution Outcome. II. Short Orbital Period Hot Subdwarf B Binaries Reveal a Clear Picture}


\author[0000-0002-6398-0195]{Hongwei Ge}
\affiliation{Yunnan Observatories, Chinese Academy of Sciences,\\
	396 YangFangWang, Guandu District, Kunming, 650216, People's Republic of China}
\affiliation{Key Laboratory for Structure and Evolution of Celestial Objects, \\
	Chinese Academy of Sciences, P.O. Box 110, Kunming 650216, People's Republic of China}
\affiliation{International Centre of Supernovae, Yunnan Key Laboratory,\\
	Kunming 650216, People's Republic of China}
\email{gehw@ynao.ac.cn}

\author[0000-0002-1556-9449]{Christopher A Tout}
\affiliation{Institute of Astronomy, The Observatories, University of Cambridge,\\
	Madingley Road, Cambridge, CB3 0HA, UK}
\email{cat@ast.cam.ac.uk}

\author{Ronald F Webbink}
\affiliation{University of Illinois at Urbana-Champaign, 1002 W Green St, Urbana, 61801, USA}
\email{rwebbink@illinois.edu}

\author[0000-0001-5284-8001]{Xuefei Chen}
\affiliation{Yunnan Observatories, Chinese Academy of Sciences,\\
	396 YangFangWang, Guandu District, Kunming, 650216, People's Republic of China}
\affiliation{Key Laboratory for Structure and Evolution of Celestial Objects, \\
	Chinese Academy of Sciences, P.O. Box 110, Kunming 650216, People's Republic of China}
\affiliation{International Centre of Supernovae, Yunnan Key Laboratory,\\
		Kunming 650216, People's Republic of China}

\author[0000-0002-1455-2784]{Arnab Sarkar}
\affiliation{Institute of Astronomy, The Observatories, University of Cambridge,\\
	Madingley Road, Cambridge, CB3 0HA, UK}

\author[0000-0002-2577-1990]{Jiao Li}
\affiliation{Key Lab of Space Astronomy and Technology, National Astronomical Observatories, \\
		Chinese Academy of Sciences, Beijing 100101, People's Republic of China}

\author[0000-0002-1421-4427]{Zhenwei Li}
\affiliation{Yunnan Observatories, Chinese Academy of Sciences,\\
	396 YangFangWang, Guandu District, Kunming, 650216, People's Republic of China}
\affiliation{Key Laboratory for Structure and Evolution of Celestial Objects, \\
	Chinese Academy of Sciences, P.O. Box 110, Kunming 650216, People's Republic of China}
\affiliation{International Centre of Supernovae, Yunnan Key Laboratory,\\
	Kunming 650216, People's Republic of China}
		
\author[0009-0001-3638-3133]{Lifu Zhang}
\affiliation{Yunnan Observatories, Chinese Academy of Sciences,\\
	396 YangFangWang, Guandu District, Kunming, 650216, People's Republic of China}
\affiliation{Key Laboratory for Structure and Evolution of Celestial Objects, \\
	Chinese Academy of Sciences, P.O. Box 110, Kunming 650216, People's Republic of China}
\affiliation{University of Chinese Academy of Sciences, Beijing 100049, People's Republic of China}

\author[0000-0001-9204-7778]{Zhanwen Han}
\affiliation{Yunnan Observatories, Chinese Academy of Sciences,\\
	396 YangFangWang, Guandu District, Kunming, 650216, People's Republic of China}
\affiliation{Key Laboratory for Structure and Evolution of Celestial Objects, \\
	Chinese Academy of Sciences, P.O. Box 110, Kunming 650216, People's Republic of China}
\affiliation{International Centre of Supernovae, Yunnan Key Laboratory,\\
		Kunming 650216, People's Republic of China}
\affiliation{University of Chinese Academy of Sciences, Beijing 100049, People's Republic of China}
\email{zhanwenhan@ynao.ac.cn}

\begin{abstract}

Common envelope evolution (CEE) is vital to form short orbital period compact binaries. It covers many objects, such as double compact merging binaries, type Ia supernovae progenitors, binary pulsars and X-ray binaries. Knowledge of the common envelope (CE) ejection efficiency still needs to be improved, though progress has been made recently. Short orbital period hot subdwarf B star (sdB) plus white dwarf (WD) binaries are the most straightforward samples with which to constrain CEE physics. We apply the known orbital period--white dwarf mass relation to constrain the sdB progenitors of seven sdB+WD binaries with a known inclination angle. The average CE efficiency parameter is 0.32. This is consistent with previous studies. However, the CE efficiency need not be constant but a function of the initial mass ratio based on well-constrained sdB progenitor mass and evolutionary stage. Our results can be used as physical inputs for binary population synthesis simulations of related objects. A similar method can also be applied to study other short orbital period WD binaries. 

\end{abstract}


\section{Introduction} 
\label{sec:intro}

The common envelope (CE) phase is fundamental to binary star evolution. The common envelope evolution (CEE) is vital in the formation of short orbital period binaries of one or two compact objects \citep{1988ApJ...329..764L,2000ARA&A..38..113T,2013A&ARv..21...59I}. These systems cover a broad range of progenitor masses and types, such as merging black-hole--black-hole, black-hole--neutron-star, and neutron-star--neutron-star binaries \citep[e.g.,][]{2021MNRAS.508.5028B,2021A&A...651A.100O,2021ApJ...920...81S,2023ApJ...955..133G,2023MNRAS.524..426I}, type Ibn/Icn supernovae from merging Wolf-Rayet/black hole \citep{2022ApJ...932...84M}, type IIb supernovae \citep{2017MNRAS.470L.102S}, luminous red novae \citep[e.g.,][]{2022ApJ...938....5M,2022A&A...667A...4C}, type Ia supernovae progenitors \citep[e.g.,][]{2018RAA....18...49W,2011MNRAS.417.1466K}, binary pulsars \citep[e.g.,][]{2013MNRAS.432L..75C,2018MNRAS.477..384L}, X-ray binaries \citep[e.g.,][]{2000ApJ...530L..93T,2000ApJ...529..946P}, double white dwarfs \citep[e.g.,][]{2023A&A...669A..82L}, cataclysmic variables \citep[e.g.,][]{1984ApJ...277..355W}, and planetary nebulae with binary nuclei \citep[e.g.,][]{1988ApJ...329..764L}. For a comprehensive review, we refer the reader to \citet{2013A&ARv..21...59I}, \citet{2023LRCA....9....2R} or \citet{2023pbse.book.....T}.

A standard method to constrain the CEE outcome is the energy formalism \citep{1984ApJ...277..355W,1984ApJS...54..335I,1988ApJ...329..764L}. However, the quantitative parameters in binary mass-transfer physics, such as CE efficiency $\alpha_\mathrm{CE}$ remain poorly studied. For example, the merging rate of double neutron stars can vary by two orders of magnitude for different CE efficiency \citep{2002ApJ...572..407B}. We can use different types of observed post-CE binaries to constrain the CEE physics. Benefiting from the well-known formation channel of short orbital-period hot subdwarf binaries, we use these objects to constrain the CEE outcome.

Hot subdwarf stars of spectral types B and O (sdB/Os) are located around the bluest end of the horizontal branch and roughly between the main sequence (MS) and white dwarf (WD) sequence in the Hertzsprung--Russell diagram \citep[reviewed by][]{2009ARA&A..47..211H,2016PASP..128h2001H}. Their surface temperatures and surface gravities are relatively high, i.e., $20,000\,\mathrm{K} \le T_\mathrm{eff} \le 70,000\,\mathrm{K}$ and $5.0 \le \log_{10}\mathrm{(g/cm\,s^{-2})} \le  6.5$ \citep{2016PASP..128h2001H,2023ApJ...942..109L}. Based on a catalogue of 39,800 hot subluminous star candidates, over one thousand sdB/Os have been identified with the Large Sky Area Multi-Object Fiber Spectroscopic Telescope spectra \citep[][]{2021ApJS..256...28L,2023ApJ...942..109L}. \citet{2020A&A...635A.193G} compiled an updated catalogue of 5874 hot subdwarf stars. Most of the sdBs are core helium-burning stars of about $0.5\,M_\sun$, with a thin envelope, on the extreme horizontal branch. About half of the sdBs reside in close binary systems \citep{2001MNRAS.326.1391M,2011MNRAS.415.1381C}. About one-third of sdB binaries have a MS companion with a long orbital period of a few hundred days \citep{2018MNRAS.473..693V}. In addition, about two-thirds of sdB binaries have a WD, a typical M-type low-mass MS star or a brown dwarf (BD) companion with short orbital periods from about 1 hr to 30 d \citep{2015A&A...576A..44K,2022A&A...666A.182S,2023A-A...673A..90S}.

Short orbital  period sdB binaries with WD companions can constrain the CE efficiency because they likely form after a first stable mass transfer phase and a second unstable CE phase \citep{2003MNRAS.341..669H,2008ASPC..392...15P}. After the first mass transfer phase, there is a well-known orbital period--WD mass ($P_\mathrm{orb}$--$M_\mathrm{WD}$) relation \citep{1971A&A....13..367R}. Hence, the sdB progenitor mass and evolutionary stage can be well determined through the WD mass in an observed sdB+WD system. We calculate the binding energy by tracing the precise change of donor's total energy as a function of its remnant mass \citep{2010ApJ...717..724G,2022ApJ...933..137G}. 

Section 2 introduces the formation channel of short orbital period sdBs with WD companions and the $P_\mathrm{orb}$--$M_\mathrm{WD}$ relation. Section 3 describes our method to calculate the binding energy and determine the CE efficiency parameter by constraining the sdB progenitor's mass and evolutionary stages. We present our results of the CE efficiency from the energy and angular momentum descriptions in Section 4. We discuss the uncertainties and make a comparison with previous studies in Section 5, 6 and 7. Section 8 summarizes and concludes our knowledge of CEE from short orbital period sdB+WD binaries.

\section{Hot Subdwarf B Star with a White Dwarf companion}

Radial velocity surveys of sdBs have provided orbital parameters of such binary systems \citep[e.g.,][]{2001MNRAS.326.1391M,2011A&A...530A..28G,2011MNRAS.415.1381C}. With this benefit, the number of short-period sdB binaries with known orbits has increased rapidly to over 140 \citep{2015A&A...576A..44K,2015MNRAS.450.3514K}. Also, \citet{2022A&A...666A.182S,2023A-A...673A..90S} derive the nature of the primaries and secondaries of more sdB binaries by classifying the space-based light-curve variations and combining these with a fit to the spectral energy distribution, the distance derived from Gaia, and the atmospheric parameters. Among these short orbital period sdB binaries \citep{2015A&A...576A..44K,2023A-A...673A..90S} over half contain a WD companion. At the same time about a quarter are accompaniedly a MS/BD star. For the rest, the type of the sdB's companions is still undecided between a MS/BD or a WD. 

\begin{figure}[htb!]
	\centering
	\includegraphics[scale=0.35]{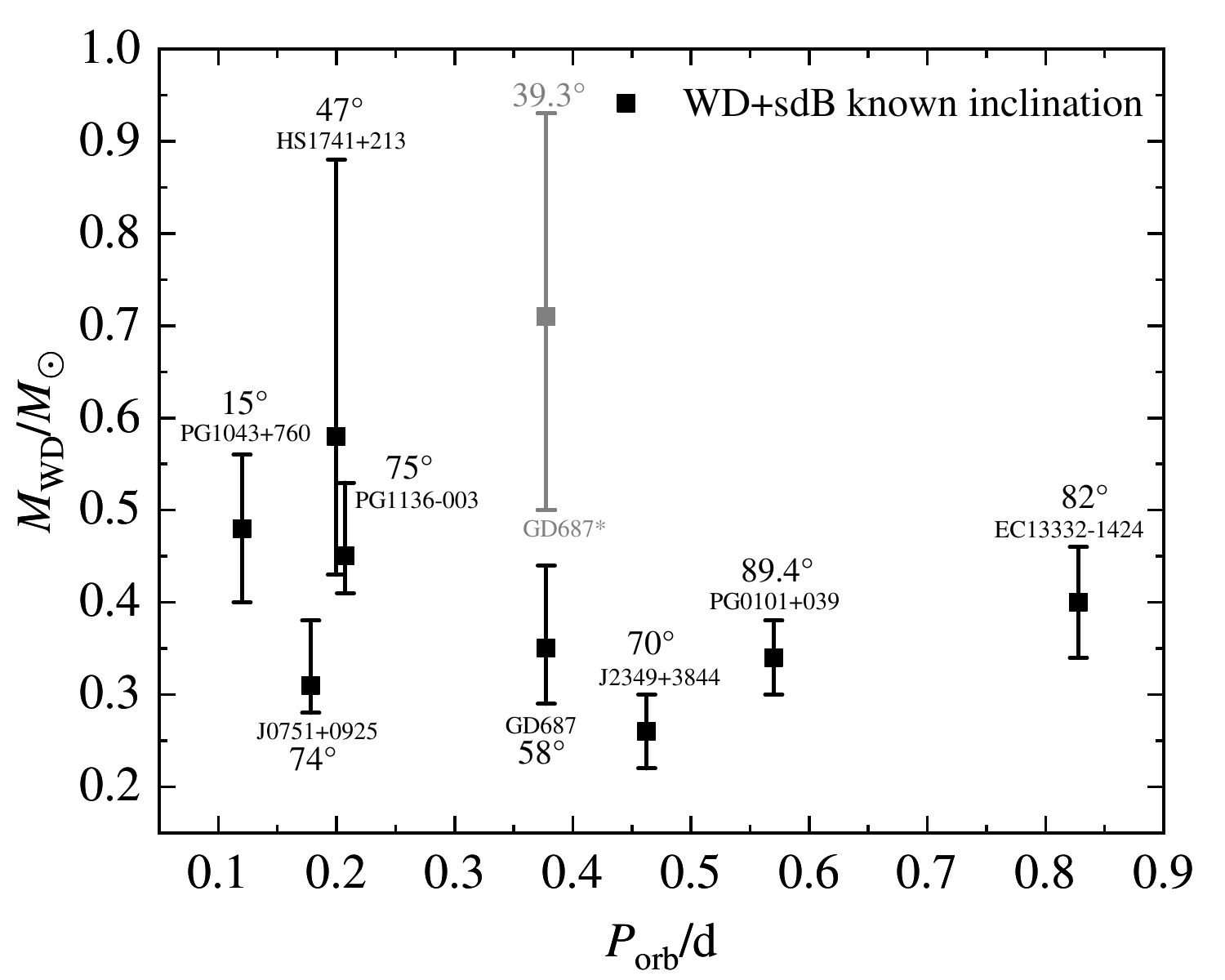}
	\caption{The observed orbital periods and white dwarf masses of 8 hot subdwarf B and white dwarf binaries with known inclination angles \citep{2023A-A...673A..90S}. The gray line marks the observed data of \citet{2010A-A...515A..37G}. 
	\label{fig1-PM1}}
\end{figure}

To precisely constrain the CE evolution physics, we need to have accurate masses of the sdB and its companion. The masses of two companions can be derived from the well-known binary mass function, 
\begin{equation}
	{f_{\mathrm{m}}=\frac{M_{\mathrm{comp}}^{3} \sin ^{3} i}{\left(M_{\mathrm{comp}}+M_{\mathrm{sdB}}\right)^{2}}=\frac{P_{\mathrm{orb}} K_1^{3}}{2 \pi G}},
	\label{fm}
\end{equation}
where $P_{\mathrm{orb}}$ is the orbital period and $K_1$ is the amplitude of the radial velocity variation. The sdB mass $M_{\mathrm{sdB}}$ and the companion mass $M_{\mathrm{comp}}$ are found when the inclination angle $i$ is known. Typically, we expect orbital inclinations to the line of sight to be uniformly distributed over a sphere. So, the distribution of $\cos i$ is flat. The minimum mass for the companion can be determined if we assume an inclination $i = \pi/2$. For a large sample of sdB binaries, we can research the inclination angel distribution and statistical features. However, we focus on precisely constraining the CE ejection process. So, we analyze short orbital sdB binaries with known inclinations \citep[see Figure~\ref{fig1-PM1} and Table~\ref{tab:sdBWD1} with data partially from][]{2023A-A...673A..90S}. The orbital parameters (period $P_{\mathrm{orb}}$, system velocity $\Gamma$, radial velocity semi-amplitude $K_1$, inclination angle $i$, and separation $A$) as well as the WD mass $M_{\mathrm{WD}}$, sdB mass $M_{\mathrm{sdB}}$, effective temperature $T_\mathrm{eff}$ and surface gravity $\log g$ are listed in Table~\ref{tab:sdBWD1}.

\begin{deluxetable*}{llrrllllll}   
	\tabletypesize{\footnotesize}
	\tablewidth{0pt}
	\tablecolumns{10}
	\tablecaption{Observed short orbital period sdB binaires with a white-dwarf companion \label{tab:sdBWD1}}
	\tablehead{
		\colhead{Name1} & \colhead{$P_\mathrm{orb}/\mathrm{d}$} & \colhead{$\Gamma/\mathrm{km\,s^{-1}}$} & \colhead{$K_1/\mathrm{km\,s^{-1}}$}
		& \colhead{$i$}  & \colhead{$A/{R_\odot}$}  & \colhead{$M_\mathrm{WD}/{M_\odot}$}
	    & \colhead{$M_\mathrm{sdB}/{M_\odot}$} & \colhead{$T_\mathrm{eff}/\mathrm{K}$} & \colhead{$\log_{10} (\mathrm{g}/\mathrm{cm\,s^{-2}})$} } 
	\startdata
	PG1043+760	&	$0.12015$	&	$24.8$	&	$63.6$	&	$15.0\pm 0.6$	&	$0.94\pm 0.04$	&	$0.48^{+0.08}_{-0.08}$	&	$0.291$	&	$27600\pm 800$	&	$5.39\pm 0.10$	\\
	GALEXJ075147.0+092526	&	$0.17832$	&	$15.5$	&	$147.7$	&	$74.0\pm 10.0$	&	$1.19\pm 0.08$	&	$0.31^{+0.07}_{-0.03}$	&	$0.365$	&	$30620\pm 490$	&	$5.74\pm 0.12$	\\
	HS1741+213	&	$0.20000$	&	\nodata	&	$157.0$	& $47.0\pm 11.0$	&	$1.40\pm 0.30$	&	$0.58^{+0.30}_{-0.15}$	&	$0.400$	&	$35600$	&	$5.30$	\\
	PG1136-003	&	$0.20754$	&	$23.3$	&	$162.0$	&	$75.0\pm 11.0$	&	$1.40\pm 0.10$	&	$0.45^{+0.08}_{-0.04}$	&	$0.501$	&	$31200\pm 600$	&	$5.54\pm 0.09$	\\
	GD687	&	$0.37765$	&	$32.3$	&	$118.3$	&	$58.0\pm 8.0$	&	$1.90\pm 0.20$	&	$0.35^{+0.09}_{-0.06}$	&	$0.283$	&	$24300\pm 500$	&$	5.32\pm 0.07$	\\
	GD687*	&	$0.37765$	&	$32.3$	&	$118.3$	&	$39.3\pm 6.0$	&	$2.32\pm 0.20$	&	$0.71^{+0.22}_{-0.21}$	&	$0.470$	&	$24300\pm 500$	&	$5.32\pm 0.07$	\\
	GALEXJ234947.7+384440	&	$0.46252$	&	$2.0$	&	$87.9$	&	$70.0\pm 10.0$	&	$2.20\pm 0.20$	&	$0.26^{+0.04}_{-0.04}$	&	$0.406$	&	$28400\pm 00$	&	$5.40\pm 0.30$	\\
	PG0101+039	&	$0.56990$	&	$7.3$	&	$104.7$	&	$89.4\pm 0.6$	&	$2.53\pm 0.01$	&	$0.34^{+0.04}_{-0.04}$	&  	$0.416$	&	$32200$	&	$5.79$	\\
	EC13332-1424	&	$0.82794$	&	$-53.2$	&	$104.1$	&	$82.0\pm 2.0$	&	$3.40\pm 0.20$	&	$0.40^{+0.06}_{-0.06}$	&	$0.400$	&	\nodata	&	\nodata	\\*
	\enddata 
	\tablecomments{Objects and data are mostly from partial results of \citet[][and references therein]{2023A-A...673A..90S} while $*$ marks that of \citet{2010A-A...515A..37G}.}
\end{deluxetable*}

Unlike the short orbital sdB + MS binaries, most short orbital sdB + WD binaries can help us narrow down the progenitor mass of a sdB with the benefits from their first mass-transfer process. Consequently, observed short orbital sdB + WD binaries help us constrain the CE physics more precisely.

\begin{figure*}[htb!]
	\centering
	\includegraphics[scale=0.545]{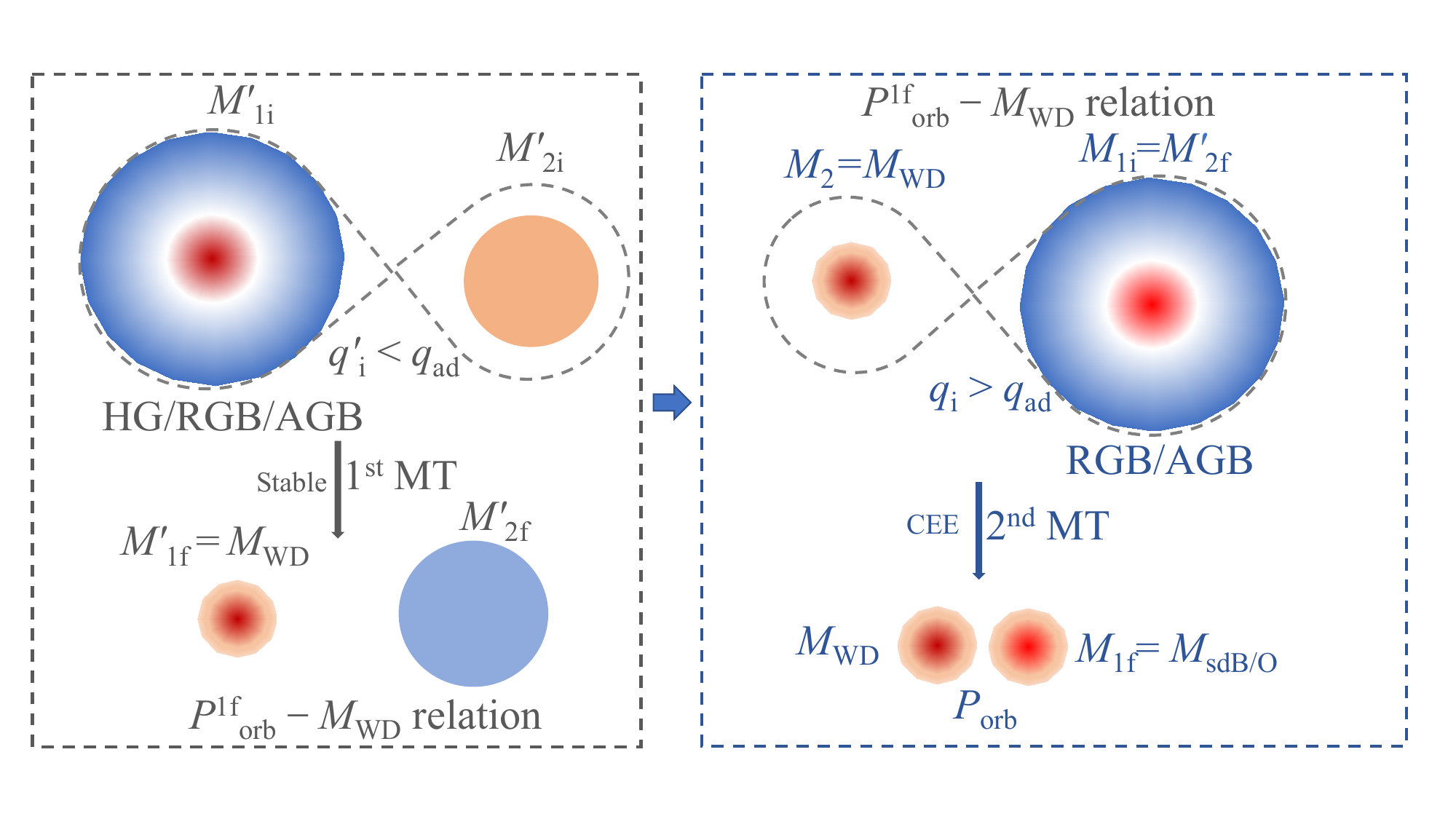}
	\caption{A typical formation channel for a close hot subdwarf B/O star (sdB/O) plus white dwarf (WD) binary. The orbital period and WD mass relation after the first stable mass transfer process consequently impact the pairing of the WD mass and the secondary mass $M_2$. This is because the secondary radius around the tip of the red giant branch is a function of its mass $M_2$ (see the second solid line from top in Figure\,\ref{fig3-MRMC}). The period-white dwarf mass relation is consistent to some degree with the distribution of the progenitors of sdB binaries before the second Common Envelope (CE) phase in the separation--secondary mass plane \citep[figure 7 of][]{2003MNRAS.341..669H}.
		\label{fig2-2MT}}
\end{figure*}

\subsection{First Stable Mass Transfer Matters}
\label{subsec:1st-MT}

\citet{2016PASP..128h2001H} reviewed the formation channels of sdB binaries in detail. The three main scenarios were proposed more than forty years ago. These are the CE evolution \citep{1976IAUS...73...75P}, stable mass transfer (also known as Roche-lobe overflow), and merging WDs \citep{1984ApJ...277..355W}. Isolated sdB stars might be formed by the merging of two helium WDs \citep{2000MNRAS.313..671S,2012MNRAS.419..452Z} or that of a red-giant core with a low-mass star, a brown dwarf or a planet \citep{1998AJ....116.1308S,2008ApJ...687L..99P}. Long orbital period sdB binaries, from 10 to over 1000\,d, are likely formed when a red-giant donor loses its envelope through dynamically stable mass transfer to a MS companion \citep{2003MNRAS.341..669H,2008ASPC..392...15P,2013MNRAS.434..186C}. Short orbital period sdB binaries, from about 1\,hr to 30\,d, undergo at least one CE phase in principle. If the companion is a MS star, the sdB binary suffers only one CE evolution. 

On the other hand, a short orbital sdB binary with a WD companion is likely formed after a first stable mass transfer phase and a second unstable CE phase \citep{2003MNRAS.341..669H,2008ASPC..392...15P}. Notably, a critical mass ratio for instability $q_\mathrm{crit} \equiv M_{\rm donor}/M_{\rm accretor} = 1.2\,{\rm or}\,1.5$ is typically adopted, independent of the progenitor mass of an sdB. Recent studies by \citet{2020ApJ...899..132G,2023ApJ...945....7G,2020ApJS..249....9G} show that the critical mass ratios for dynamically unstable $q_\mathrm{ad}$ or thermal timescale mass transfer $q_\mathrm{th}$ depend on both the progenitor mass and evolutionary stage (the radius).

We need to know the progenitor mass, initial mass ratio, and initial period before the CE as well as the final remnant mass, final mass ratio and final orbital period after the CE to fit the CE evolution precisely (right panel of Figure\,\ref{fig2-2MT}). For short orbital period sdB binaries with WD companions, the current masses of both companions and orbital period or separation after the second CE evolution are known. The progenitor mass of sdB before the second CE evolution can be constrained by considering the first stable mass transfer phase (left panel of Figure\,\ref{fig2-2MT}). After this phase, the $P_\mathrm{orb}$--$M_\mathrm{WD}$ relation means the WD and progenitor mass do not pair arbitrarily. Because the progenitors of sdBs are almost all located at the tip of the red giant branch (TRGB), where radii are a function of mass.

\subsection{Period--White Dwarf Mass Relation}
\label{subsec:PM}

\begin{figure}[htb!]
	\centering
	\includegraphics[scale=0.44]{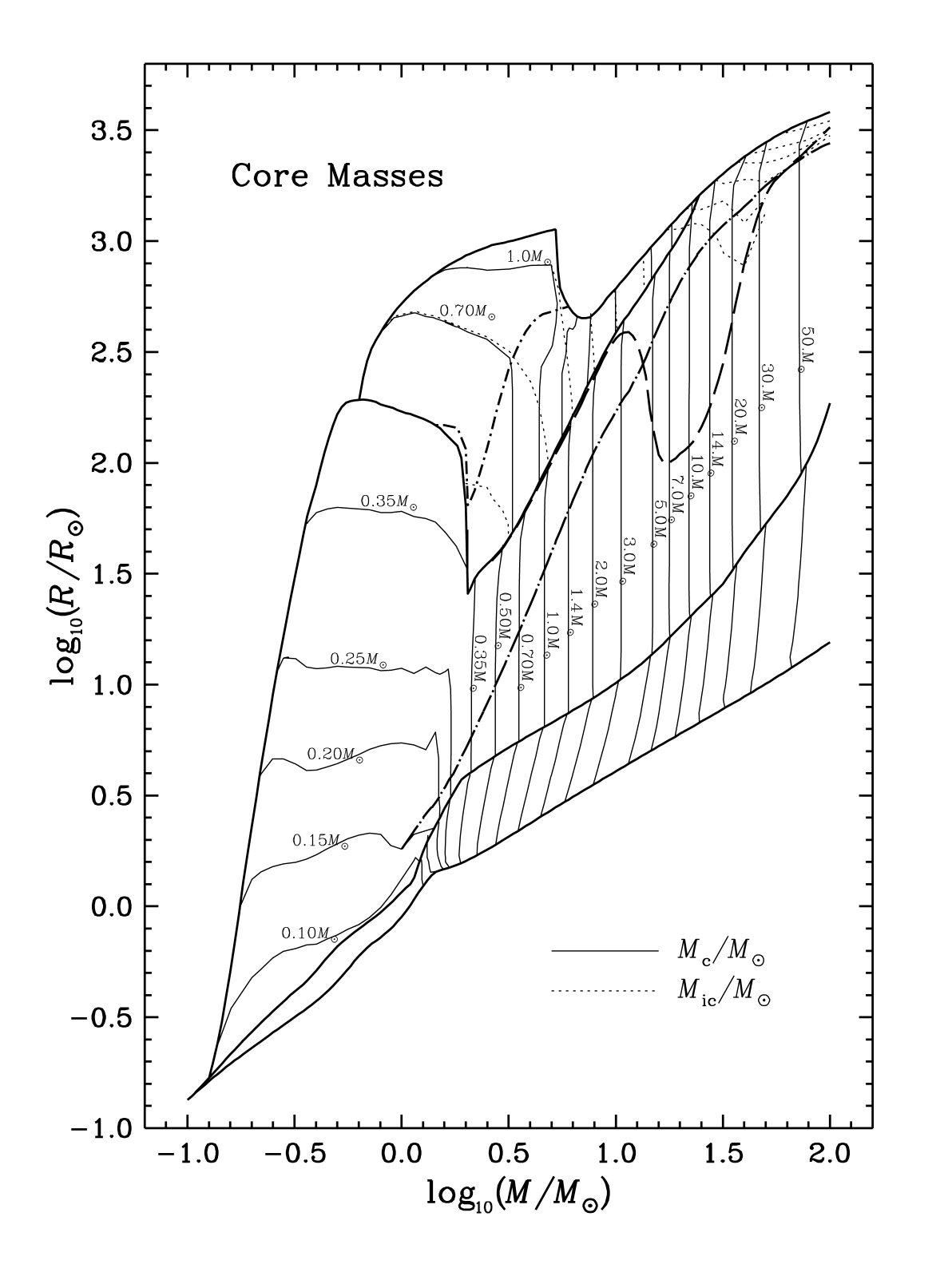}
	\caption{Core masses $M_\mathrm{c}$ and inner core masses $M_\mathrm{ic}$ on the mass--radius diagram. Stellar models cover a mass range from 0.1 to 100\,$M_\sun$ with metallicity $Z=0.02$ (\citealt{2020ApJ...899..132G}). The thick solid lines mark crucial evolutionary stages, the zero-age main sequence (ZAMS), the terminal main sequence (TMS), the tip of the red giant branch (BGB) and the tip of the asymptotic giant branch (TAGB). The dash-dotted line labels the base of red giant branch (BGB). The thin-solid and thin-dotted lines show all stars' core and inner core masses at different evolutionary stages. The masses, $M_\mathrm{c}$ and $M_\mathrm{ic}$ mark the midpoints in hydrogen and helium depletion profiles, respectively. In other words, the core mass $M_\mathrm{c}$ refers to the mass coordinate at which the helium abundance is halfway between the surface helium abundance and the maximum helium abundance in the stellar interior. The inner core mass $M_\mathrm{ic}$ identifies the mass coordinate at which the helium abundance is halfway between the maximum helium abundance in the stellar interior and the minimum helium abundance interior to that maximum.
		\label{fig3-MRMC}}
\end{figure}

\citet{1971A&A....13..367R} found a relation between helium WD mass $M_\mathrm{WD}$ and orbital separation $A$ through slow and conservative Case B \citep{1967ZA.....65..251K} mass transfer for a set of $2.5M_\sun$ total-mass binaries. The first reason comes from the well-known core mass--radius ($M_\mathrm{c}$--$R$) relation for giants \citep{1971AcA....21..417P,1983ApJ...270..678W,1987ApJ...319..180J}. This $M_\mathrm{c}$--$R$ relation derives from the luminosity $L$ depending only on the core mass $M_\mathrm{c}$ and the radius $R$ depending only on the luminosity $L$ and total mass $M_\mathrm{d}$ for donor stars with degenerate cores. We illustrate the core mass $M_\mathrm{c}$ and inner core mass of $Z=0.02$ donor stars on the mass and radius diagram in Figure\,\ref{fig3-MRMC}. Stellar model grids in this figure are collected from detailed numerical simulations by \citep{2020ApJ...899..132G}. The second reason stems from mass-transfer physics. The Roche-lobe radius of the donor star, for mass ratio $0<M_\mathrm{d}/M_\mathrm{a}<0.8$, can be approximated by
\begin{equation}
	R_\mathrm{L}=\frac{2}{3^{4 / 3}} A\left(\frac{M_\mathrm{d}}{M_\mathrm{d}+M_\mathrm{a}}\right)^{1 / 3}
\end{equation}
\citep{1971ARA&A...9..183P}, where $M_\mathrm{a}$ is the mass of the accretor. From Kepler's law, we have
\begin{equation}
	\left(\frac{2 \pi}{P_\mathrm{orb}}\right)^2 A^3=G\left(M_\mathrm{d}+M_\mathrm{a}\right),
\end{equation}
where $P_\mathrm{orb}$ is the orbital period and $G$ is Newton's gravitational constant. For a binary system undergoing stable mass transfer (Roche-lobe overflow, RLOF), the donor radius $R$ almost closely tracks its Roche-lobe radius $R_\mathrm{L}$. Combining the above knowledge from mass transfer physics, we find that the total mass of the donor and accretor is eliminated. \citet{1995MNRAS.273..731R} derived the expression for the orbital period 
\begin{equation}
	P_\mathrm{orb}=20\,G^{-1 / 2} R^{3 / 2} M_\mathrm{d}^{-1 / 2}.
	\label{Porb}
\end{equation}
The core mass--radius relation can be used in Equation (\ref{Porb}), and the remnant mass $M_\mathrm{WD}$ of a donor with a degenerate core is close to the core mass $M_\mathrm{c}$. So, after stable mass transfer stops, we expect a relation between the orbital period $P_\mathrm{orb}$ and the donor's remnant mass $M_\mathrm{WD}$.

\citet[][hereinafter RPJDH95]{1995MNRAS.273..731R} studied the detailed mass transfer process in a binary system, initially containing a neutron star and a low-mass giant and finally ending up as a wide binary comprising a radio pulsar and a WD \cite[see, e.g.,][]{1991PhR...203....1B}. RPJDH95 devised a fitting formulae for $P_\mathrm{orb}$--$M_\mathrm{WD}$ relations, 
\begin{equation}
\frac{P_{\mathrm{orb}}}{\rm d} = 0.374\left[ \frac{\left( \frac{R_{0}}{R_\odot}\right)  \left(\frac{M_\mathrm{WD}}{M_\odot}\right)^{4.5}} {\left(1+4 \left(\frac{M_{\mathrm{WD}}}{M_\odot}\right)^{4}\right)}+0.5\right]^{3 / 2} \left(\frac{M_{\mathrm{WD}}}{M_\odot}\right)^{-1 / 2}.
\label{RP95}
\end{equation}
The above relations cover core mass range $0.15<M_\mathrm{WD}/M_\sun<1.15$. The radii $R_0$ depend on the metallicity $Z$. For $Z=0.02$ stellar models, $Z=0.001$ models, and all combined models, $R_0$ equals $5500\,R_\sun$, $3300\,R_\sun$, and $4950\,R_\sun$, respectively.

Using an updated Cambridge STARS code \citep{1995MNRAS.274..964P}, \citet[][hereinafter TS99]{1999A&A...350..928T} derived new $P_\mathrm{orb}$--$M_\mathrm{WD}$ correlations based on 121 LMXB models. The best fit for all models is
\begin{equation}
{\frac{M_\mathrm{WD}}{M_{\odot}}=\left(\frac{(P_{\mathrm{orb}}/{\rm d})}{b}\right)^{1 / a}+c},
\label{TS99}
\end{equation}
where, depending on the chemical composition of the donor,
\begin{equation}
	(a, b, c)=\left\{\begin{array}{llll}
		(4.50, & 1.2 \times 10^{5}, & 0.120) &   Z=0.02 \\
		(4.75, & 1.1 \times 10^{5}, & 0.115) & \text{Combined} \\
		(5.00, & 1.0 \times 10^{5}, & 0.110) & Z=0.001.
	\end{array}\right.
\end{equation}

\citet[][hereinafter CHDP13]{2013MNRAS.434..186C} presented a systematic study \citep[extended by][]{2021MNRAS.502..383Z} of the long-period sdB binaries. They found that sdB binaries produced from stable mass transfer follow a unique $P_\mathrm{orb}$--$M_\mathrm{WD}$ relation for progenitor masses below the helium flash limits. The remnant mass in this study covers a relatively massive range from 0.37 to 0.51 $M_\sun$. \citet{2014MNRAS.437.2217S} examined binary millisecond pulsars that form after a stable mass transfer phase from a low to intermediate mass companion, between central hydrogen exhaustion and core helium ignition, to a neutron star. They confirmed that their stellar models reproduced a well-defined $P_\mathrm{orb}$--$M_\mathrm{WD}$ relation,
\begin{equation}
\frac{P_\mathrm{orb}}{\mathrm{d}} =\exp \left( a + \left( \frac{b}{M_\mathrm{WD}/M_\odot}\right) + c \left( \ln \frac{M_\mathrm{WD}}{M_\odot} \right) \right),
\label{ST14}
\end{equation}
with
\begin{equation}
{\begin{array}{l}
	(a, b, c) \\
	\,=\left\{\begin{array}{ll}
			(22.902,3.097,23.483) &  0.170 \leq M_{\mathrm{WD}}<0.225 \\
			(14.443,0.335,9.481) &  0.225 \leq M_{\mathrm{WD}}<0.280 \\
			(11.842,-3.831,-4.146) &  0.280 \leq M_{\mathrm{WD}}<0.480.
		\end{array}\right.
\end{array}}
\end{equation}

\begin{figure}[htb!]
	\centering
	\includegraphics[scale=0.35]{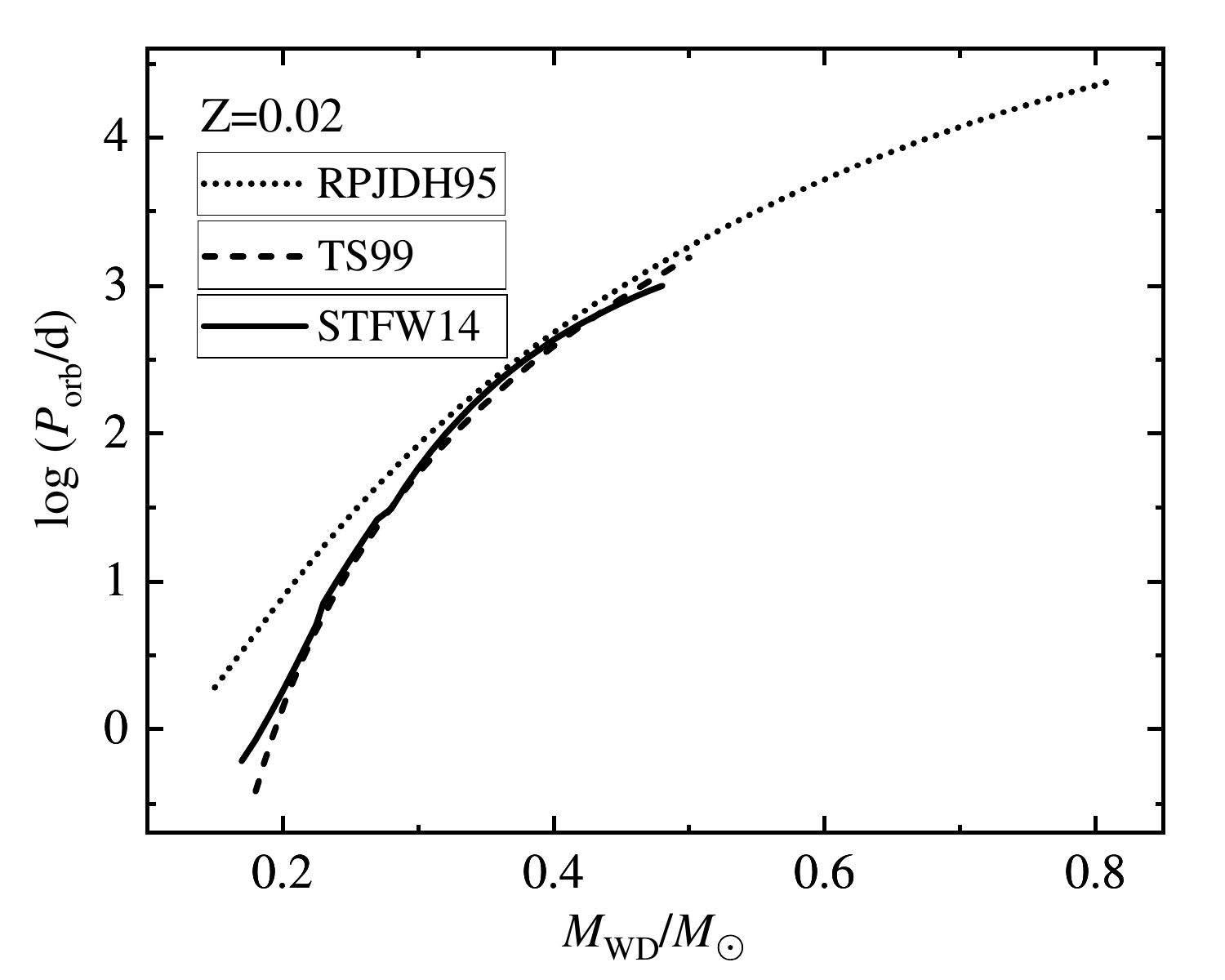}
	\caption{The period--WD mass ($P_\mathrm{orb}$--$M_\mathrm{WD}$) relation shows in the works of literature (metallicity $Z=0.02$). The dotted line, dashed line, and solid line represent the fitting formulae by \citet[][RPJDH95]{1995MNRAS.273..731R}, \citet[][TS99]{1999A&A...350..928T}, and \citet[][STFW14]{2014MNRAS.437.2217S}, respectively. We cut the WD mass off at around $0.8\,M_\odot$ because of the upper limit of the minimum WD masses \citep{2022A&A...666A.182S}.
	\label{fig4-PM}}
\end{figure}
The $P_\mathrm{orb}$--$M_\mathrm{WD}$ relations from different studies share a similar slope. However, the $P_\mathrm{orb}$--$M_\mathrm{WD}$ relation difference becomes significant for relatively less massive ($M_\mathrm{WD} < 0.35\,M_\odot$) WDs (see Figure~\ref{fig4-PM}). For Pop II stars ($Z=0.001$), the orbital period is systematically smaller than that for $Z=0.02$ stars with the same WD mass (RPJDH95, TS99, and STFW14). Different chemical abundances and mixing length parameters can change the orbital period in the $P_\mathrm{orb}$--$M_\mathrm{WD}$ relation by a factor of 2.4 (RPJDH95). It should be noticed first that the orbital period is independent of the accretor mass (all previous studies and Equation \ref{Porb}). Furthermore the $P_\mathrm{orb}$--$M_\mathrm{WD}$ relation is not much affected by the non-conservative nature of stable mass transfer (TS99, CHDP13, and STFW14) and so the fraction of mass or angular momentum loss from the binary system. However, for an individual binary, TS99 pointed out that $P_\mathrm{orb}$ and $M_\mathrm{WD}$ increase with increasing fractions of mass and angular momentum loss. \citet{2023MNRAS.525.2605G} studied the $P_\mathrm{orb}$--$M_\mathrm{WD}$ relation under wind mass-loss and pointed out that tidelly enhanced wind can play an important role for helium WD binaries in wide orbits. In addition, the $P_\mathrm{orb}$--$M_\mathrm{WD}$ relation is ascertained from a circular orbit. For some binary pulsars with a possible neutron star or massive WD ($M> 0.8M_\odot$), the eccentricity is high ($e>0.025$, see RPJDH95) . Such high eccentricity is not the case for progenitors of sdB+WD binaries and it is safe to assume a circular orbit in this study. Finally, in this study, we mainly focus on low-mass donors with $M_\mathrm{d} < 2.0\,M_\sun$ and so the $P_\mathrm{orb}$ and $M_\mathrm{WD}$ relation fits well for these stars. However, there is a slight deviation of the $P_\mathrm{orb}$ and $M_\mathrm{WD}$ relation for more massive donors (see TS99 and Figure 7 of \citealt{2003MNRAS.341..669H}). TS99 explained this because the mass transfer becomes dynamically unstable for $M_\mathrm{d} > 2.0\,M_\sun$ stars. A systematical study of the criteria for dynamical timescale mass transfer by \citet{2020ApJ...899..132G} shows that $6.3\,M_\sun > M_\mathrm{d} > 1.8\,M_\sun$ giant stars more easily enter dynamically unstable mass transfer than less massive giants (see Figures 6 and 7 of \citealt{2020ApJ...899..132G}).

\section{Methods}

We study the CE ejection process under the assumption that it is adiabatic. This might not be entirely appropriate for the whole ejection process. However, it works to constrain the initial and final state of CE evolution. After the hydrodynamical interaction, the remnant star could relax to hydrostatic and subsequent thermal equilibrium. The adiabatic mass loss model and the code have been introduced in detail by \citet[][hereinafter Paper I]{2022ApJ...933..137G}. The physical inputs and control parameters, such as opacity tables and the mixing length parameter, are described in Paper I and references therein. 

We follow a standard procedure to predict the outcome of CE evolution. This is based on the essential physics of total energy conservation, known as the energy formalism \citep{1984ApJ...277..355W,1984ApJS...54..335I,1988ApJ...329..764L}. It is written as
\begin{equation}
\alpha_\mathrm{CE} \Delta E_\mathrm{orb} = E_\mathrm{bind},
\label{alpha1}
\end{equation}
where $\alpha_\mathrm{CE}$ is the CE efficiency parameter, the fraction of the orbital energy change $\Delta E_\mathrm{orb}$ used to overcome the envelope's binding energy $E_\mathrm{bind}$.

After a dimensionless parameter $\lambda$ is introduced in the binding energy term, the formula is often written as \citep{1984ApJ...277..355W,1990ApJ...358..189D}
\begin{equation}
	\alpha_{\mathrm{CE}}\left(\frac{G M_{1 \mathrm{i}} M_{2}}{2 A_{\mathrm{i}}} - \frac{G M_{1 \mathrm{f}} M_{2}}{2 A_{\mathrm{f}}}\right)=-\frac{G M_{1 \mathrm{i}} M_{1 \mathrm{e}}}{\lambda R_{1 \mathrm{i}}},
\label{lambda}
\end{equation}
where $G$ is Newton’s gravitational constant, $M_\mathrm{1i}$ and $M_\mathrm{1f}$ are the initial and final mass of the donor, $M_2$ is the accretor's mass assumed not to change, $A_\mathrm{i}$ and $A_\mathrm{f}$ are the initial and final semimajor axes, $M_\mathrm{1e} = M_\mathrm{1i} - M_\mathrm{1f}$ is the envelope mass of the donor, $R_\mathrm{1i}$ is the initial radius of the donor and $\lambda$ is a dimensionless parameter that reflects the structure of the donor star. This $\lambda$ was previously assumed to be 0.5 in binary population synthesis studies. However, many recent works \citep{2001A&A...369..170T,2010ApJ...716..114X,2011ApJ...743...49L,2011MNRAS.411.2277D} aimed precisely calculate it and significantly improved our knowledge of the binding energy of stars with different masses and evolutionary stages. 

The binding energy can be calculated for the initial giant as
\begin{equation}
		\begin{aligned}
		E^{'}_\mathrm{bind} & =\int_{M_\mathrm{c}}^{M_{\mathrm{li}}}\left(-\frac{G m}{r}+\alpha_\mathrm{th}U\right) d m,\\
		& =E_\mathrm{grav} + \alpha_\mathrm{th} E_\mathrm{th}
	\end{aligned}
\label{Ebind1}
\end{equation}
where $\alpha_\mathrm{th}$ is defined as the fractional contribution of the internal energy \citep{1995MNRAS.272..800H} or thermal efficiency. We assume $\alpha_\mathrm{th}=1$ throughout this paper unless specific notes are made. The core response, after the CE ejection, is neglected in this expression. This is fine for donor stars with degenerate cores. However, it can cause problems for massive stars with non-degenerate cores. In particular, companions of massive stars with different masses could spiral into different depths inside the non-degenerate core. As described in Paper I, we calculate the binding energy $E_\mathrm{bind}$ by tracing the change of the total energy of the donor as the mass is lost \citep{2010ApJ...717..724G},
\begin{equation}
	\begin{aligned}
		  \Delta E_1 \equiv -E_{\mathrm{bind}} = & \int_{0}^{M_{\mathrm{lf}}}\left(-\frac{G m}{r}+U\right) d m \\
		& -\int_{0}^{M_{\mathrm{li}}}\left(-\frac{G m}{r}+U\right) d m,
\end{aligned}
\label{Ebind2}
\end{equation}
where $m$ is the donor mass, $r$ is the radius and $U$ is its specific internal energy including recombination part. The final mass $M_{\mathrm{lf}}$ is not fixed to the core mass $M_{\mathrm{c}}$ as usual but determined by when the remnant first shrinks back inside its Roche lobe $R_\mathrm{1f} < R_\mathrm{1,L1}$. Our method considers the redistribution of the remnant core with a thin envelope and the impact of the companion mass. In addition, we also compare the companion's radius $R_2$ to its Roche lobe radius $R_\mathrm{2,L1}$. So, we have
\begin{equation}
 R_{\mathrm{1f}} \le R_{\mathrm{1,L1}}, {\rm and} ~
 R_{\mathrm{2}} \le R_{\mathrm{2,L1}},
	\label{radius}
\end{equation}
for a successive CE ejection. If $R_\mathrm{1f} \le R_\mathrm{1,L1}$ and $R_2 \le R_\mathrm{2,L1}$, the envelope is, in principle, ejectable. Otherwise, the CE evolution ends with merging if $R_\mathrm{1f} > R_\mathrm{1,L1}$ and $R_2 > R_\mathrm{2,L1}$. Aside from the role of the donor, the companion type could play an important role in the outcome of CE evolution. The MS companion radius may be larger than its Roche-lobe radius if its mass is too close to the donor mass. So, a non-compact companion with mass close to its donor may cause the stars to merge in CE evolution. However, the WD companion radius is almost always smaller than its Roche-lobe radius.

\subsection{The Initial to Final Separation Relation in Different Formulisms}

With the calculation of binding energy in Equation\,\ref{Ebind2}, we define the CE efficiency parameter as $\beta_{\rm CE}$ to distinguish from the binding energy calculation in Equation\,\ref{lambda}. The final to initial separation relation can be written as
\begin{equation}
	{\frac{A_{\mathrm{f}}}{A_{\mathrm{i}}}=\frac{M_{1 \mathrm{f}}}{M_{1 \mathrm{i}}}\left(1+\frac{2 A_{\mathrm{i}} \Delta E_{1}}{\beta_{\mathrm{CE}} G M_{2} M_{1 \mathrm{i}}}\right)^{-1} .}
	\label{beta}
\end{equation}

We apply our constraint to CE evolution using short orbital period sdB plus WD binaries and present our results in the next section. We compare our results to recent studies by \citet{2022MNRAS.517.2867H}, \citet{2022MNRAS.513.3587Z} and \citet{2023MNRAS.518.3966S} on WD binaries with low-mass main-sequence, white dwarf companions or brown dwarf companions. So, we introduce here other formulae that predict the outcome of CE evolution from its initial conditions that differ from the energy conservation description. 

\citet{2023ApJ...944...87D} use angular momentum in a new way to derive a simple expression for the final orbital separation and their method is well suited to higher-order multiples. We adopt their expressions for the $\eta$ formalism \citep{2023ApJ...944...87D}, the $\gamma$ formalism \citep{2000A&A...360.1011N,2005MNRAS.356..753N}, and the $\alpha$ formalism \citep{1984ApJ...277..355W,1990ApJ...358..189D}. The $\eta$ formalism is written as
\begin{equation}
	\begin{aligned}
	\frac{A_\mathrm{f}}{A_\mathrm{i}}= & \left[\frac{(1+q)\left(1+q_{\mathrm{ec}}\right)^{2}}{1+q\left(1+q_{\mathrm{ec}}\right)}\right] \\
	& \times \exp \left[-2 \eta\left(\frac{q\, q_{\mathrm{ec}}}{1+q}\right) \mathcal{F}(q, \delta)\right],
	\end{aligned}	
	\label{eta}
\end{equation}
where
\begin{equation}
\mathcal{F}(q, \delta)=\mathcal{Q}(q, \delta) \frac{1}{q}+\mathcal{Q}\left(\frac{1}{q}, \delta\right) q,
\end{equation}
and
\begin{equation}
	\mathcal{Q}(q, \delta) \equiv \frac{[f(q)]^{\delta}}{[f(q)]^{\delta}+\left[f\left(q^{-1}\right)\right]^{\delta}} .
\end{equation}
The function of $f(q)$ is defined by \citet{1983ApJ...268..368E} as
\begin{equation}
	f(q) = r_\mathrm{L}(q)= \frac{R_\mathrm{L}}{A} =\frac{0.49 q^{\frac{2}{3}}}{0.6 q^{\frac{2}{3}}+\ln \left(1+q^{\frac{1}{3}}\right)},
\end{equation}
and the mass ratios $q \equiv M_\mathrm{1f}/M_\mathrm{2}$ while $q_{\mathrm{ec}} \equiv M_\mathrm{1e}/M_\mathrm{1f} $. They suggest using $\delta = 3$ to constrain the post-CE binaries.
The $\alpha$ formalism can be rewritten as
\begin{equation}
	\frac{A_\mathrm{i}}{A_\mathrm{f}}=\left(1+q_{\mathrm{ec}}\right)\left(1+\frac{2 q_{\mathrm{ec}}\, q}{\alpha_\mathrm{CE} \lambda}\left[f\left(q+q_{\mathrm{ec}} q\right)\right]^{-1}\right),
	\label{alpha2}
\end{equation}
and the $\gamma$ formalism for
\begin{equation}
	\begin{aligned}
		\frac{A_\mathrm{i}}{A_\mathrm{f}}= & \left[\frac{1}{1+q_{\mathrm{ec}}}\right]^{2}\left[\frac{1+q\left(1+q_{\mathrm{ec}}\right)}{1+q}\right] \\
		&\times\left[\frac{1+q\left(1+q_{\mathrm{ec}}\right)}{1+q+q\,q_{\mathrm{ec}}(1-\gamma)}\right]^{2}.
		\label{gamma}
	\end{aligned}
\end{equation}

\subsection{The Model Grids for the Progenitors of Hot Subdwarf B Stars}

\begin{figure}[htb!]
	\centering
	\includegraphics[scale=0.35]{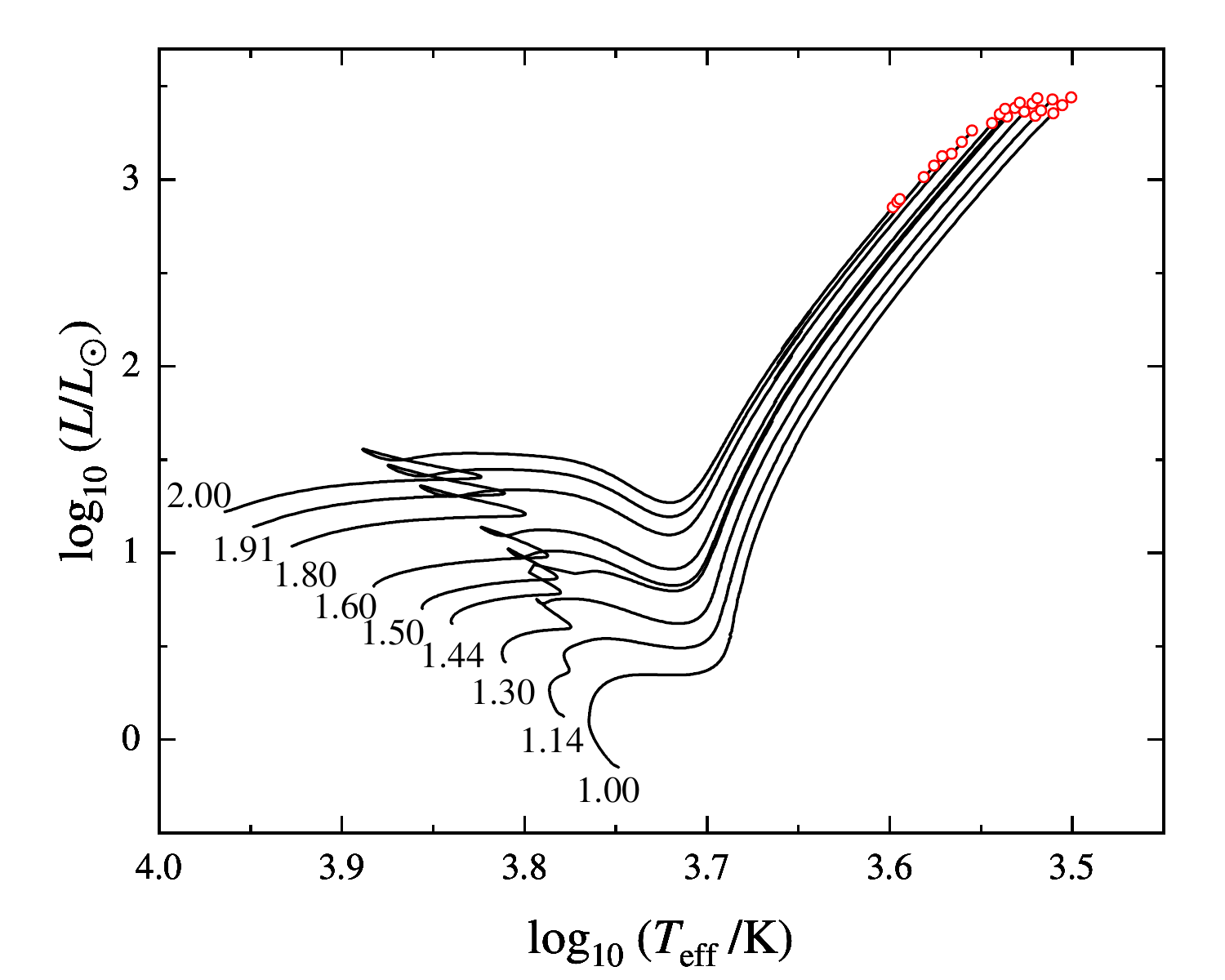}
	\caption{Model grids for sdB progenitors are marked with red open circles. The evolutionary stages in a narrow region near the TRGB with masses from 1.00 to 2.00 $M_\odot$ and metallicity $Z=0.02$.
		\label{fig-1to2}}
\end{figure}

In Paper I, we aim to preliminarily check the precisely calculated binding energy of which the response of the donor remnant and the effects of companion mass are naturally considered. However, the progenitor masses of the observed sdBs are not strictly constrained. We apply the $P_\mathrm{orb}$--$M_\mathrm{WD}$ relation to constrain the progenitor mass for sdB+WD binaries. The progenitor model grids are extended as shown in Figure\,\ref{fig-1to2}. The mass grids are expanded to 1.00, 1.14, 1.30, 1.44, 1.50, 1.60, 1.80, 1.91, and 2.00 $M_\odot$. The evolutionary stages are covered in a narrow region \citep{2002MNRAS.336..449H,2021MNRAS.505.3514Z} near the TRGB (open circles in Figure\,\ref{fig-1to2}). We ignore progenitors with a mass larger than $2\,M_\odot$ because the $P_\mathrm{orb}$--$M_\mathrm{WD}$ relation is likely unsuitable for RGB stars with non-degenerate cores. The influence of this should be negligible in our study because most of the progenitor masses of sdBs are likely around 1.0 to 1.3 $M_\odot$.

With the model grids, we trace the radius response and the total energy change (Equation~\ref{Ebind2}) of the donor using the adiabatic mass-loss code \citep{2010ApJ...717..724G,2010ApSS.329..243G,2015ApJ...812...40G,2020ApJ...899..132G,2023ApJ...945....7G}. We first assume the progenitor masses of eight sdB binaries with a WD companion in Table\,\ref{tab:sdBWD1} are 1.00, 1.14, 1.30, 1.44, 1.50, 1.60, 1.80, 1.91, and 2.00 $M_\odot$. Starting with an initial guess of $\beta_\mathrm{CE} = 0.5$ we use a bisection method to solve Equations \ref{Ebind2}, \ref{radius}, \ref{beta} and \ref{fm}. According to the difference between the derived orbital period and observed orbital period, we adjust the theoretical CE efficiency parameter. Then, we solve for the CE efficiency parameter $\beta_\mathrm{CE}$ that satisfies the sdB mass $M_\mathrm{sdB}$, the WD companion mass $M_\mathrm{WD}$, the inclination angle $i$, the amplitude $K_1$, and the orbital period $P_\mathrm{orb}$. Secondly we constrain the $\beta_\mathrm{CE}$ range using the error bar of the observed WD mass. 

Finally, we restrict the possible $\beta_\mathrm{CE}$ by determining the progenitor mass of sdB via the $P_\mathrm{orb}$--$M_\mathrm{WD}$ relation. With known $M_\mathrm{WD}$ in an sdB + WD binary, the theoretical orbital period $P_\mathrm{theo}$ is calculated by Equation\,\ref{ST14} from STFW14. The initial separation $A_\mathrm{i}$ just before the CE can also be determined from the Roche-lobe radius of the donor
\begin{equation}
	A_\mathrm{i}  = \frac{R_\mathrm{L}}{r_\mathrm{L}(q)} = \frac{R_\mathrm{1i}}{r_\mathrm{L}(M_\mathrm{1i}/M_2)} .
\end{equation} The initial orbital period $P_\mathrm{i}$ can be derived from Kepler's third Law.

\section{Results}

\begin{figure}[htb!]
	\centering
	\includegraphics[scale=0.35]{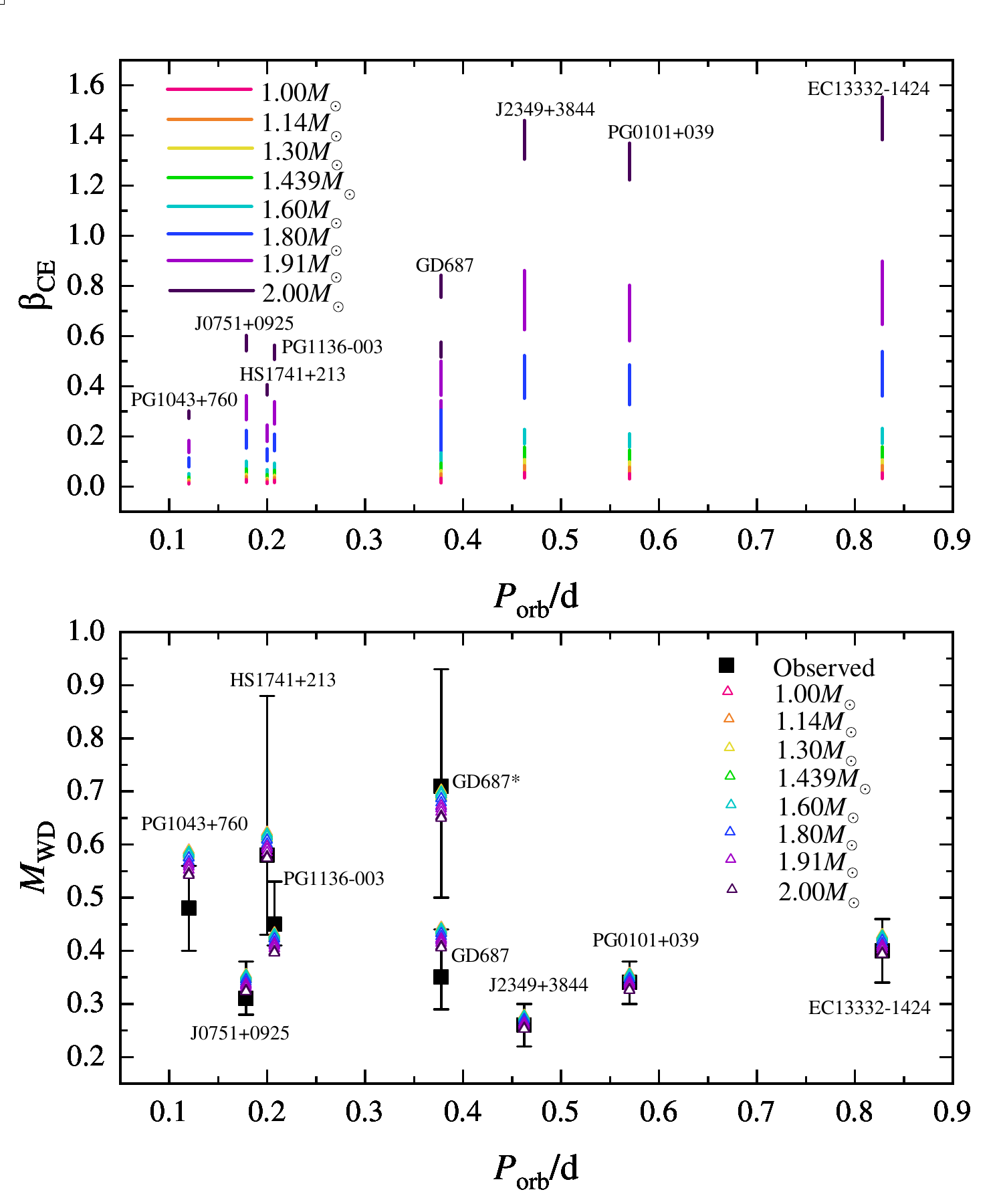}
	\caption{The CE efficiency parameter $\beta_\mathrm{CE}$ for sdB + WD binaries with progenitor masses between 1.00 and 2.00\,$M_\odot$ (top panel) and WD mass that pair the final sdB mass for the corresponding progenitor mass with the orbital period.
		\label{fig-beta-all}}
\end{figure}

We first assume the progenitors of sdBs are located near the TRGB with masses between 1.00 and 2.00\,$M_\odot$ and metallicity $Z=0.02$, as shown in Figure\,\ref{fig-1to2}. We use the method in the last section to constrain the CE efficiency parameter of 8 sdB + WD binaries in Table\,\ref{tab:sdBWD1}. The CE efficiency parameter $\beta_\mathrm{CE}$ range is relatively broad from 0.01 to 1.55 (see the top panel in Figure\,\ref{fig-beta-all}). Such results are not surprising without knowing the precise masses and evolutionary stages of the sdB progenitors. The observed WD masses alone in these sdB binaries are probably not sufficient to limit the progenitor masses of sdBs. The WD masses which pair sdB progenitors with masses between 1.00 and 2.00\,$M_\odot$ are all within the error bars except for the partially less massive progenitors for PG 1043+760 and a small region of the more massive progenitors for PG 1136-003 (see the bottom panel in Figure\,\ref{fig-beta-all}).

Aside from the WD mass $M_\mathrm{WD}$, we further apply the $P_\mathrm{orb}$--$M_\mathrm{WD}$ relation to constrain the progenitor mass of an sdB. The progenitor mass and evolutionary stage are chosen if there is a minimum of $|P_\mathrm{i} - P_\mathrm{theo}|/P_\mathrm{theo}$ from the model grids (Figure\,\ref{fig-1to2}). Then, we successfully identify the progenitor masses and radii of 7 systems at the onset of the CE process (see Table\,\ref{tab:sdBM1i}). The minima $|P_\mathrm{i} - P_\mathrm{theo}|/P_\mathrm{theo}$ for GALEXJ075147.0+092526, PG1136-003, PG0101+039, and EC13332-1424 are all less than 0.002. The minimum $|P_\mathrm{i} - P_\mathrm{theo}|/P_\mathrm{theo}$ of GD687* is 0.15. We find the $P_\mathrm{orb}$--$M_\mathrm{WD}$ relation becomes less restrictive for $M_\mathrm{WD} >0.5M_\odot$ or $M_\mathrm{WD} <0.27M_\odot$. For HS1741+213 and GALEXJ234947.7+384440, we only determine the sdB mass and radius from the WD mass $M_\mathrm{WD}$. The sdB masses of PG1043+760 and GD687 are below 0.29\,$M_\odot$ (Table\,\ref{tab:sdBWD1}). The core mass of TRGB stars decreases from about 0.47 to 0.39\,$M_\odot$ for progenitor mass from 1 to 2\,$M_\odot$. So, the progenitor masses of these two objects are likely larger than 2.00\,$M_\odot$. 

\begin{deluxetable*}{lllllllllllllll}   
	\tabletypesize{\footnotesize}
	\tablewidth{0pt}
	\tablecolumns{15}
	\tablecaption{Possible progenitors of sdBs \label{tab:sdBM1i}}
	\tablehead{
		\colhead{Name1} & \colhead{$M_\mathrm{1i}/M_\odot$} & \colhead{$R_\mathrm{1i}/R_\odot$} & \colhead{$M_\mathrm{sdB}/M_\odot$} &
		\colhead{$R_\mathrm{1f}/R_\odot$} & \colhead{$M_\mathrm{WD}/M_\odot$} & \colhead{$\frac{|dP|}{P}$} & \colhead{$\beta_\mathrm{CE}$} & \colhead{${\log_{10} (-E_\mathrm{bind}/{\rm erg})}$} & \colhead{$\lambda$}& \colhead{$q$} & \colhead{$q_\mathrm{ec}$} & \colhead{$\eta$} & \colhead{$\gamma$} & \colhead{$\alpha_\mathrm{CE} \lambda$} } 
	\startdata
    J0751+0925	&	1.91	&	79.498	&	0.410	&	0.479	&	0.333& 0.001
    	&	0.316	&	46.814	&	2.073	&	1.230	&	3.661	&	1.695	&	1.230	&	0.661	\\
    HS1741+213	&	2.00	&	60.569	&	0.392	&	0.493	&	0.578& 0.385	&	0.366	&	47.016	&	1.941	&	0.679	&	4.141	&	2.074	&	1.199	&	0.720	\\
    PG1136-003	&	1.00	&	151.460	&	0.444	&	0.539	&	0.425& 0.001	&	0.025	&	45.802	&	2.195	&	1.046	&	1.258	&	5.133	&	1.729	&	0.056	\\
    GD687*	&	1.00	&	174.780	&	0.464	&	0.795	&	0.702& 0.152	&	0.014	&	45.581	&	3.055	&	0.660	&	1.165	&	7.112	&	1.782	&	0.045	\\
    J2349+3844	&	1.91	&	79.498	&	0.411	&	0.923	&	0.260& 0.652	&	0.744	&	46.804	&	2.120	&	1.577	&	3.661	&	1.410	&	1.212	&	1.602	\\
    PG0101+039	&	1.91	&	79.498	&	0.411	&	1.040	&	0.335& 0.002	&	0.693	&	46.803	&	2.125	&	1.225	&	3.661	&	1.505	&	1.210	&	1.498	\\
    EC13332-1424	&	1.14	&	164.530	&	0.463	&	1.376	&	0.430& 0.001	&	0.053	&	45.736	&	3.267	&	1.077	&	1.472	&	3.830	&	1.589	&	0.174	\\*
	\enddata 
	\tablecomments{1. The sdB masses of PG1043+760 and GD687 are below 0.29\,$M_\odot$. Their progenitor masses are probably larger than 2.00\,$M_\odot$. \\
	2. The sdB masses and evolutionary stages of HS1741+213 and J2349+3844 are only constrained from the closest $M_\mathrm{WD}$. The $P_\mathrm{orb}$--$M_\mathrm{WD}$ relation becomes less restrictive for $M_\mathrm{WD} >0.5M_\odot$ or $M_\mathrm{WD} <0.27M_\odot$.\\
	3. The $*$ marks the data of \citet{2010A-A...515A..37G}. The minimum $|P_\mathrm{i} - P_\mathrm{theo}|/P_\mathrm{theo}$ of GD687 is 0.15. For the rest, $|P_\mathrm{i} - P_\mathrm{theo}|/P_\mathrm{theo}$ values is less than 0.002.
	}
\end{deluxetable*}

\begin{figure}[htb!]
	\centering
	\includegraphics[scale=0.35]{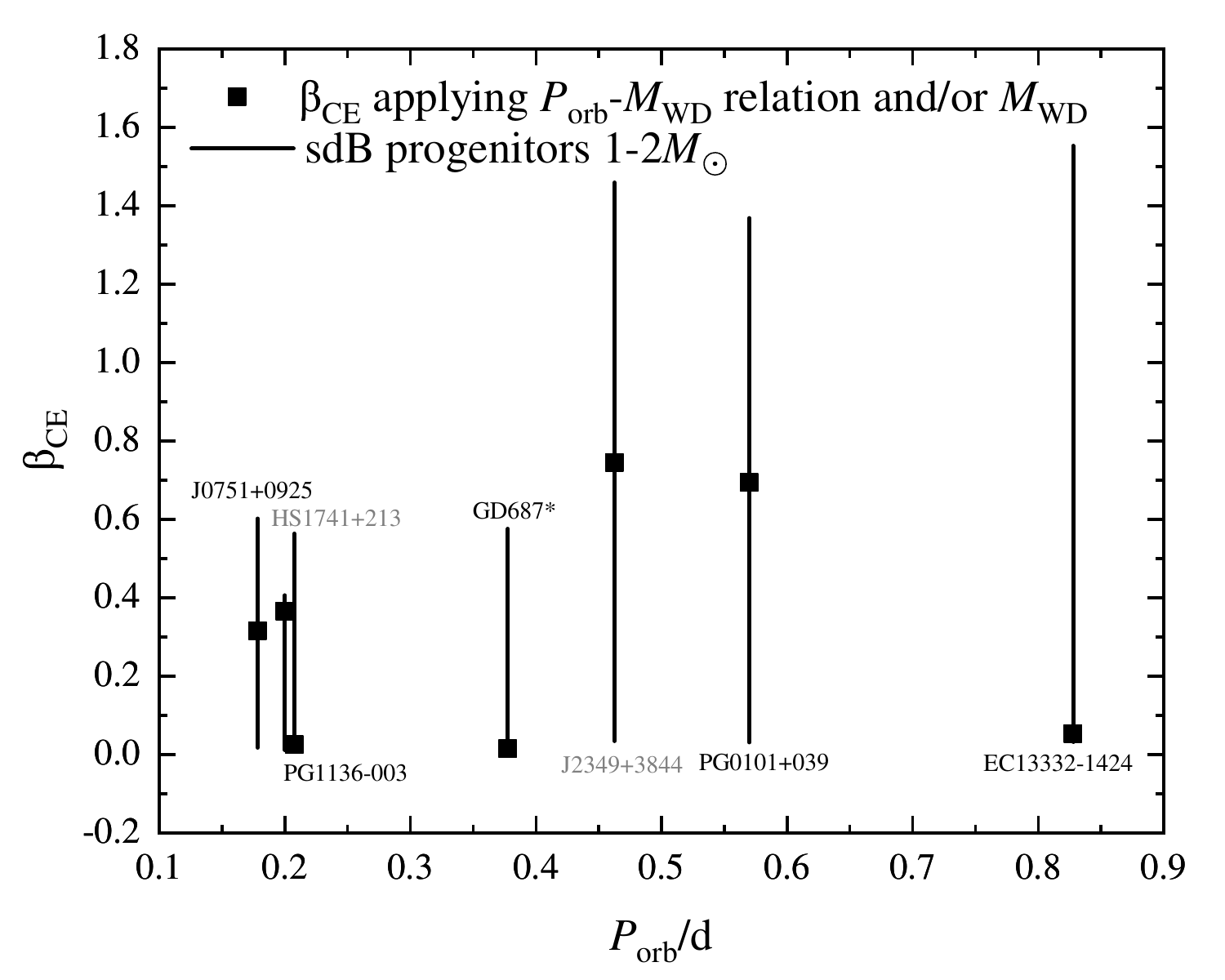}
	\caption{The CE efficiency parameter $\beta_\mathrm{CE}$ for 7 sdB+WD objects. The solid lines are for sdB progenitor masses from 1 (bottom) to 2\,$M_\odot$ (top). The filled squares are constrained by applying the $P_\mathrm{orb}$--$M_\mathrm{WD}$ relation and WD mass. Object names in gray do not follow the $P_\mathrm{orb}$--$M_\mathrm{WD}$ relation.
    \label{fig-beta-PM}}
\end{figure}

\begin{figure}[ht!]
	\centering
	\includegraphics[scale=0.35]{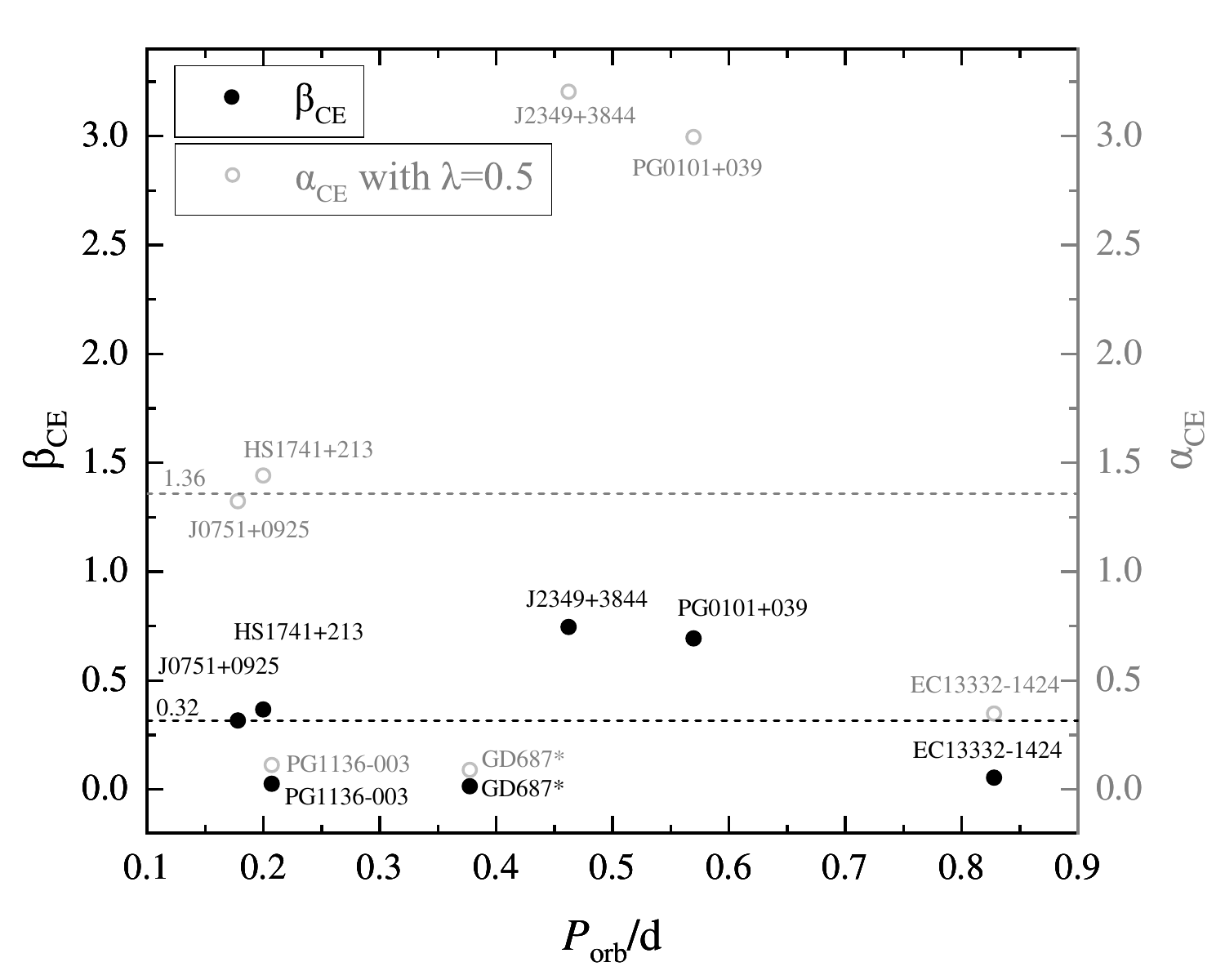}
	\caption{A comparison of CE efficiency parameters $\beta_\mathrm{CE}$ and $\alpha_\mathrm{CE}$ for sdB+WD binaries. The binding energy is calculated by Equation\,\ref{Ebind2} for $\beta_\mathrm{CE}$ (black dots) or by assuming $\lambda = 0.5$ on the right side of Equation\,\ref{lambda} (gray circles). Black and gray dashed lines show the average $\beta_\mathrm{CE}$ and $\alpha_\mathrm{CE}$. Without realistic binding energy, the CE efficiency parameter $\alpha_\mathrm{CE}$ might be inauthentic and greater than 1.
		\label{fig-beta-com}}
\end{figure}

\begin{figure}[htb!]
	\centering
	\includegraphics[scale=0.32]{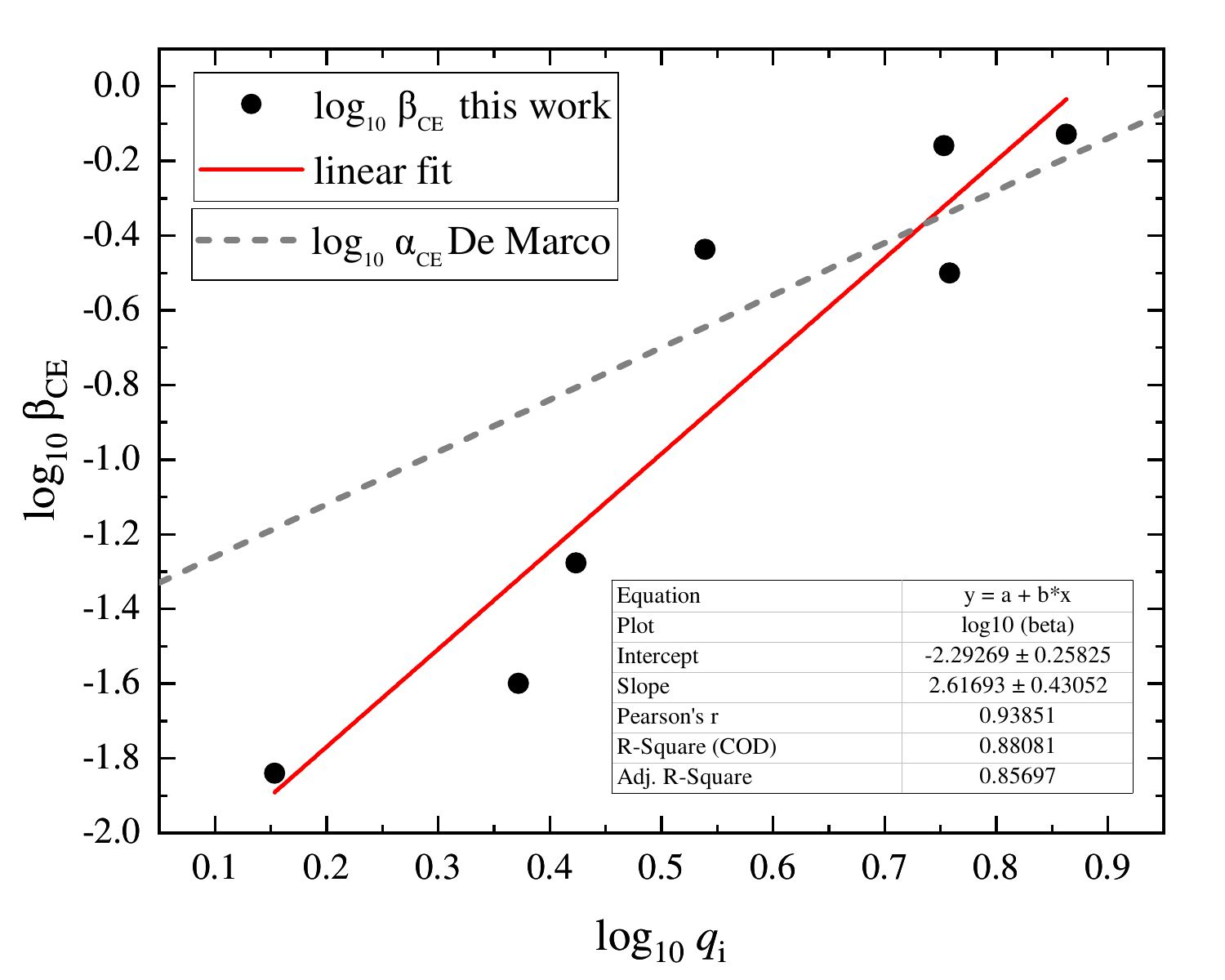}
	\caption{The linear fitting formula of the CE efficiency parameter $\log_{10} \beta_\mathrm{CE}$ as a function of the initial mass ratio $\log_{10} q_\mathrm{i} = \log_{10} M_\mathrm{1i}/M_\mathrm{WD}$ (red solid line). For comparison, we plot the fit of the CE efficiency parameter and initial mass ratio for post-red giant branch binaries and central stars of the planetary nebula by \citet{2011MNRAS.411.2277D}.
		\label{fig-beta-fit}}
\end{figure}

Once the progenitor mass and evolutionary stage at the onset of the CE phase have been determined in reverse, the CE efficiency parameter $\beta_\mathrm{CE}$ of a sdB+WD object can be determined precisely rather than in a broad range (see Figure\,\ref{fig-beta-PM}). The CE efficiency parameter $\beta_\mathrm{CE}$ is less than 0.1 for a relatively low mass progenitor. However,  $\beta_\mathrm{CE}$ is larger than 0.3 for a relatively massive progenitor (see both Figure\,\ref{fig-beta-PM} and Table\,\ref{tab:sdBM1i}). The average $\beta_\mathrm{CE}$ is 0.32 for these sdB+WD objects (Figure\,\ref{fig-beta-com}). The CE efficiency parameter $\alpha_\mathrm{CE}$ can be nonphysically larger than 3 without precisely calculated binding energy (gray circles in Figure\,\ref{fig-beta-com}). We present the linear fitting formula of the CE efficiency parameter $\beta_\mathrm{CE}$ as a function of the initial mass ratio $q_\mathrm{i} =M_\mathrm{1i}/M_\mathrm{WD}$. Figure\,\ref{fig-beta-fit} shows that the best fit is
\begin{equation}
	\log_{10} \beta_\mathrm{fit} = -2.29269 + 2.61693 \times \log_{10} q_\mathrm{i},
	\label{beta-fit1}
\end{equation}
with a variance $s^2 = \Sigma_{i=1}^N (\log_{10} \beta_\mathrm{CE}- \log_{10} \beta_\mathrm{fit})^2 /N = 0.051$. A similar relation (the gray dashed line in Figure\,\ref{fig-beta-fit}) of the CE efficiency parameter and the initial mass ratio was found by \citet{2011MNRAS.411.2277D}. The gray-dashed line represents their relation from the post-RGB stars or the central stars of the planetary nebula. In their work, the progenitor masses are from 1 to $5\,M_\odot$, while in this study, the mass range is from 1 to $2\,M_\odot$. This work's steeper gradient might indicate that the lower-mass RGB stars are more unbound than massive RGB stars.

\begin{figure*}[htb!]
 	\centering
 	\includegraphics[scale=0.6]{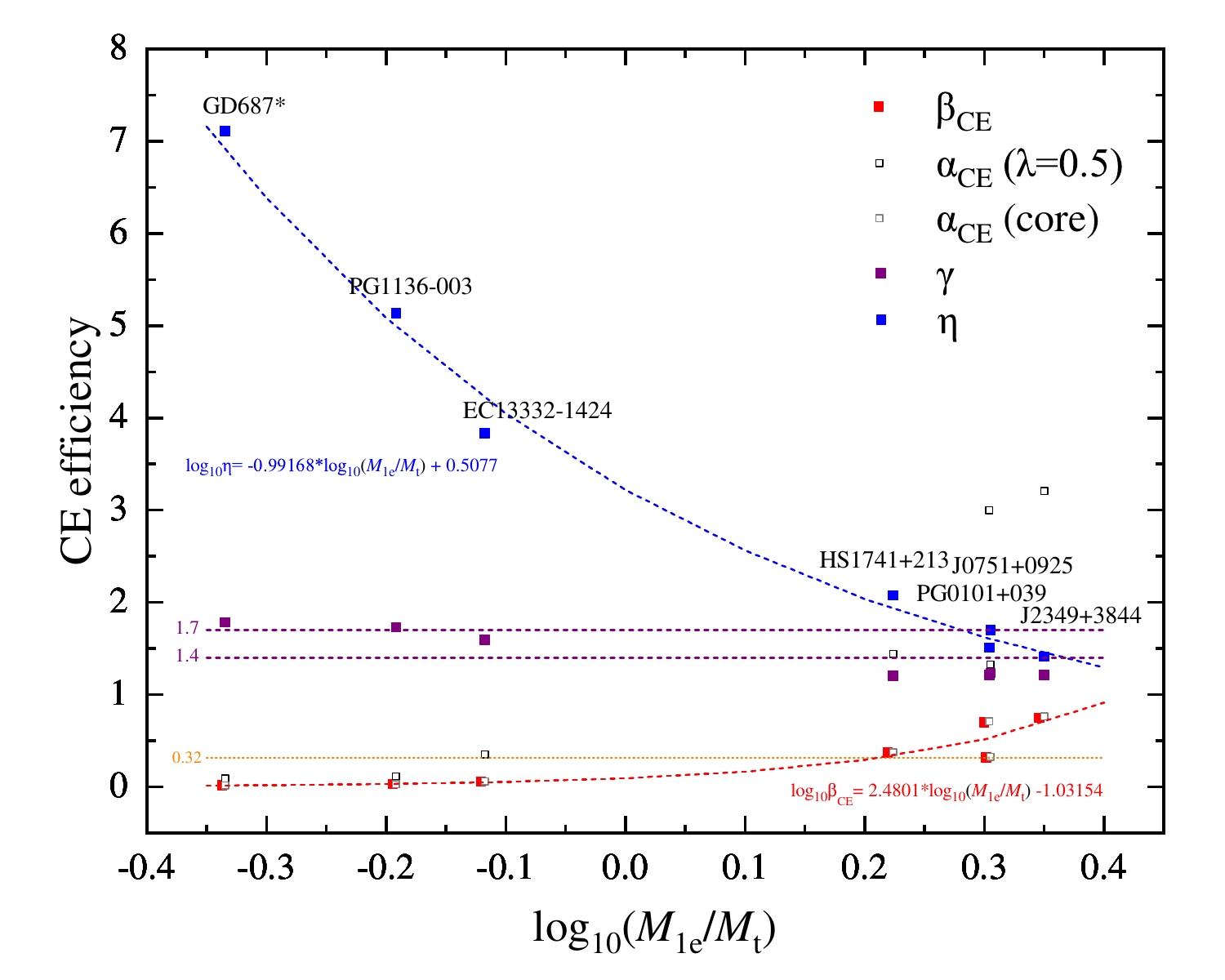}
 	\caption{A systematic comparison of parameters in different CE formalisms for 7 observed sdB+WD binaries. The x-axis is defined as $M_\mathrm{1e}/M_\mathrm{t} = (M_\mathrm{1i}-M_\mathrm{1f})/(M_\mathrm{1f}+M_\mathrm{2})$. Red squares present the $\beta_\mathrm{CE}$ values from Equation\,\ref{beta}. Black and gray open-squares show the $\alpha_\mathrm{CE}$ values from Equation\,\ref{alpha2} with the binding energy $E_\mathrm{bind}$ from $\lambda = 0.5$ and Equation\,\ref{Ebind1}. Purple squares demonstrate $\gamma$ values from Equation\,\ref{gamma}. Blue squares are $\eta$ values from Equation\,\ref{eta}.
 		\label{fig-CEE-all}}
 \end{figure*}

We systematically compare the CE efficiency parameters in different formalisms for sdB+WD binaries with determined progenitor mass and evolutionary stage. The $\alpha$ and $\beta$ formalisms are based on an energy conservation assumption, and can be calculated from Equations\,\ref{beta} and \ref{alpha1} (or \ref{alpha2}). Suppose we assume the structure parameter $\lambda = 0.5$ in Equation\,\ref{lambda}. The CE efficiency parameter $\alpha_\mathrm{CE}$ can increase to over 3\footnote{The upper limit of $\alpha_\mathrm{CE}$ in our treatment is, in principle, 1. However, some new studies claim $\alpha_\mathrm{CE}$ as large as three or more. A more significant $\alpha_\mathrm{CE}$ might mean that there is another energy source, such as jets \citep{2023MNRAS.523..221G},  absorption of nuclear luminosity \citep{2001ASPC..229..261I} or radiative pressure (private communication with Zhuo Chen).} for J2349+3844 and PG0101+039 (black open-squares in Figure\,\ref{fig-CEE-all}). However, suppose we use $E_\mathrm{bind}$ from Equation\,\ref{Ebind1} based on the core mass of the initial model. In that case, we find the CE efficiency parameter $\alpha_\mathrm{CE}$ (gray open-squares) to be almost the same as $\beta_\mathrm{CE}$ (red squares in Figure\,\ref{fig-CEE-all}). The CE efficiency parameters $\beta_\mathrm{CE}$ are derived from Equations\,\ref{Ebind2} and \ref{beta}. We fit the CE efficiency parameter $\beta_\mathrm{CE}$ as a function of the envelope mass of the donor and the sum of the remnant mass of the donor and the WD mass as
\begin{equation}
	\log_{10} \beta_\mathrm{CE} = -1.03154 + 2.4801 \times \log_{10} \frac{M_\mathrm{1e}}{M_\mathrm{1f}+M_\mathrm{WD}}.
	\label{beta-fit2}
\end{equation}

The $\gamma$ and $\eta$ formalisms are based on angular momentum, and can be calculated from Equations\,\ref{gamma} and \ref{eta}. We plot $\gamma$ and $\eta$ of these sdB+WD binaries with purple and blue squares in Figure\,\ref{fig-CEE-all}. We found $\gamma$ is between 1.20 and 1.78. Inspired by \citet{2023ApJ...944...87D}, we fit $\eta$ for seven short orbital sdB+WD binaries and find
\begin{equation}
	\log_{10} \eta = 0.5077 -0.99168 \times \log_{10} \frac{M_\mathrm{1e}}{M_\mathrm{1f}+M_\mathrm{WD}}.
\end{equation}

\section{Uncertainties in the Progenitor Mass and the Evolutionary Stage}

 Determination of the progenitor mass and the evolutionary stage becomes indispensable in understanding the detailed CE evolution. As discussed by \citet{2023MNRAS.518.3966S}, the main uncertainty in the CE efficiency parameter $\alpha_\mathrm{CE}$ arises from the uncertain progenitor mass of WD binaries. In addition, different evolutionary stages (RGB or asymptotic giant branch) of the progenitor \citep{2010A&A...520A..86Z} should also play an important role. Recently, \citet{2022MNRAS.513.3587Z} constrained the WD progenitor mass through the age of the simultaneously born brown dwarf companion. Similarly, the progenitor mass of an inner WD binary could be constrained if the age of a third star in a widely separated tertiary is known \citep{2016ComAC...3....6T}.

We use the previously found orbital period and WD mass ($P_\mathrm{orb}$--$M_\mathrm{WD}$) relation (STFW14) to constrain the sdB progenitor mass and the evolutionary stage. The CE efficiency parameter could be determined more precisely once the sdB progenitor mass and the evolutionary stage are constrained. The sdB progenitor could have accreted mass from its companion in the first stable mass transfer process, and the progenitor mass and lifetime could be slightly changed \citep{2023MNRAS.518.3966S}. The main effect would be on the lifetime and the ratio of the core mass and the envelope mass of the sdB progenitor. Our results should be the same: the sdB progenitor remains in quasi-thermal equilibrium during the first mass transfer.

In principle, the $P_\mathrm{orb}$--$M_\mathrm{WD}$ relation can constrain the mass and evolutionary state of degenerate core progenitors that have suffered a first stable mass transfer and later undergo a CE phase. It is necessary for us to comprehend the mass range of the WD and the proper metallicity before using the relation. The STFW14 $P_\mathrm{orb}$--$M_\mathrm{WD}$ relation works fine for short orbital sdB+WD binaries with $M_\mathrm{WD}$ from 0.27 to 0.50\,$M_\odot$. In addition to sdB+WD binaries, our method to constrain the progenitors by applying the $P_\mathrm{orb}$--$M_\mathrm{WD}$ relation can be extended to other WD binaries. 

\section{Uncertainties in the Binding Energy}

To constrain the CE evolution outcome accurately, we need to solve for the envelope binding energy precisely. We discuss four major impacts on this binding energy. First, we find in Figure 4 of Paper I that companions with different masses can spiral into different positions of the donor. This means that the remnant of the donor may have a slightly thicker envelope mass for a more massive companion. In this case, the absolute value of binding energy is supposed to be slightly smaller than that customarily defined at the helium core. Secondly, the donor's remnant is supposed to expand after the thick envelope is ejected. Hence, the absolute value of binding energy can be slightly smaller again. However, even though the donor remnant can expand by a factor of 2, the binding energy hardly changes for donor stars with a degenerate core. Thirdly, all the contribution of the internal energy in Equation \ref{Ebind1} is considered. The positive internal energy $E_\mathrm{th}$ plus the negative potential energy $E_\mathrm{grav}$ make the total binding energy $E_\mathrm{bind}$ less negative or even positive for relatively less massive TRGB stars \citep{2012ASPC..452....3H}. Fourth, we build the progenitor model grids without considering stellar winds. \citet{2023MNRAS.518.3966S} argue that winds may decrease the progenitor mass by up to about $10$ percent \citep{2011MNRAS.411.2277D} at the time of the CE evolution, but is a small effect relative to other uncertainties. \citet{1988MNRAS.231..823T} would argue for even more mass loss prior to CE evolution.

\begin{figure}[htb!]
	\centering
	\includegraphics[scale=0.35]{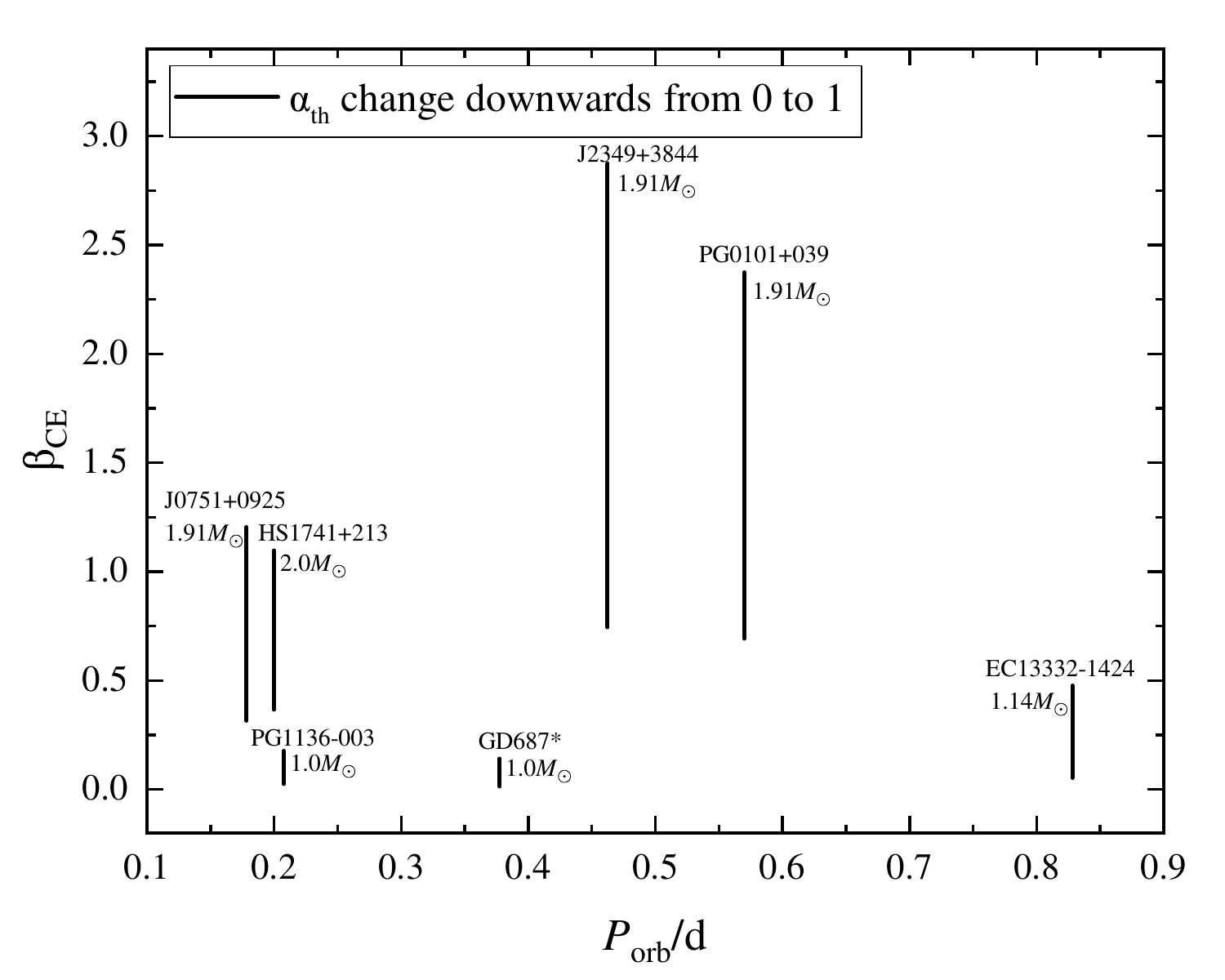}
	\caption{The impact of the internal energy on the CE efficiency parameter. 
		\label{fig-beta-Eint}}
\end{figure}

\begin{figure}[htb!]
	\centering
	\includegraphics[scale=0.35]{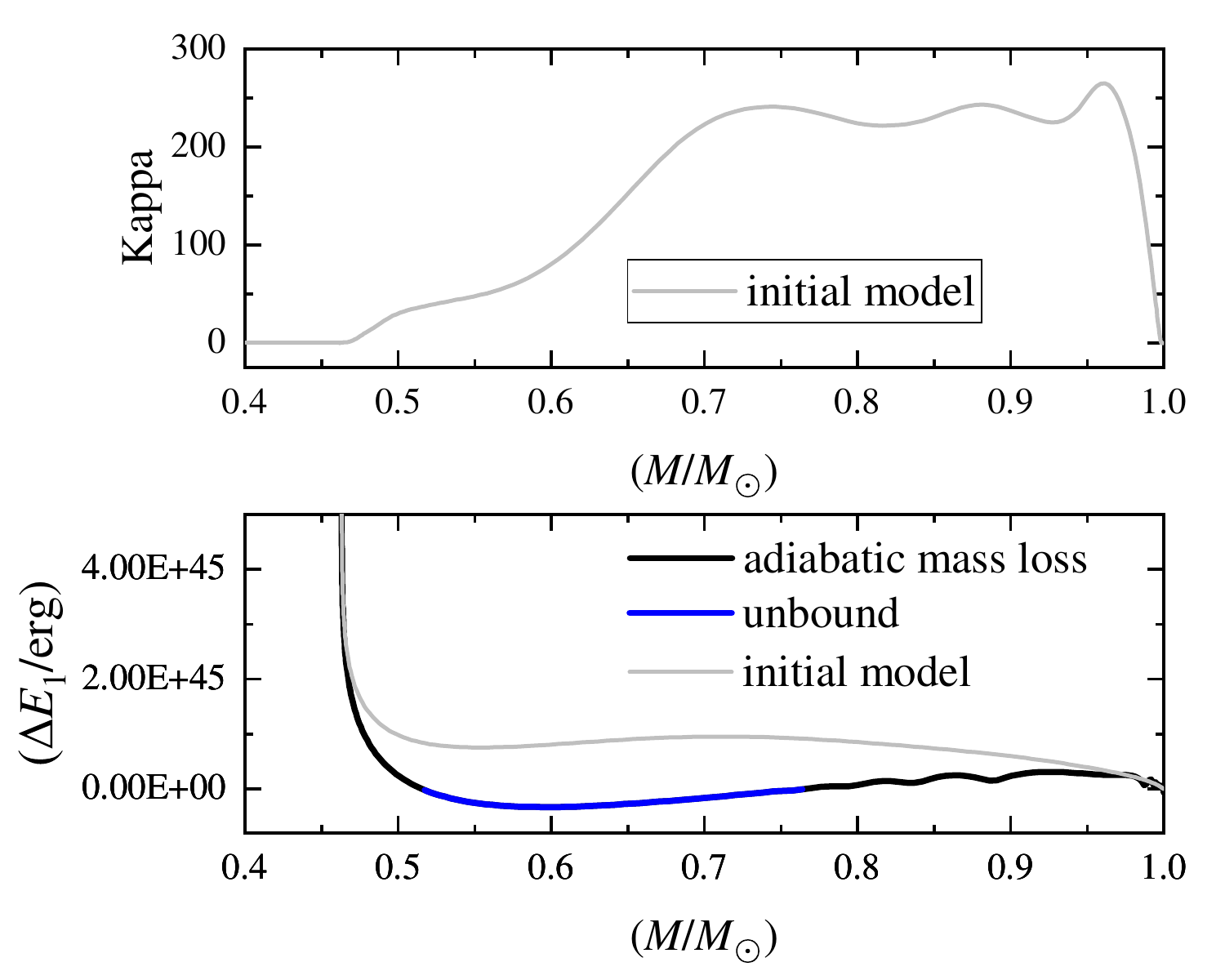}
	\caption{The opacity profile (upper panel) and different $E_\mathrm{bind}$ against the mass or remant mass of the $1\,M_\odot$ and $174.78\,R_\odot$ TRGB star.
		\label{fig-Ebind}}
\end{figure}

We examine the influence of the thermal efficiency $\alpha_\mathrm{th}$ on the CE efficiency parameter $\beta_\mathrm{CE}$. By assuming $\alpha_\mathrm{th}$ decreases from 1 to 0, the contribution of the internal energy diminishes and vanishes. The binding energy $E_\mathrm{bind}$ becomes more negative and corresponds to a more significant CE efficiency parameter $\beta_\mathrm{CE}$ (Figure\,\ref{fig-beta-Eint}). Theoretically, the maximum of $\beta_\mathrm{CE}$ should be 1 in our treatment. However, if the companion launches jets, the CE efficiency parameter can be larger \citep{2023MNRAS.523..221G}. So, for relatively more massive progenitors ($1.9$ or $2.0M_\odot$), the thermal efficiency $\alpha_\mathrm{th}$ is suggested to be close to 1 (Figure\,\ref{fig-beta-Eint}). If there is a constant CE efficiency parameter, the thermal efficiency of relatively less massive progenitors is likely to be smaller. 

As we discussed in Paper I, we calculate the binding energy of the donor by tracing the change of total energy as a function of its remnant mass (see Equation,.\ref{Ebind2}). So, the first three impacts on the binding energy are naturally addressed. Compared with the initial model before CE, the envelope response becomes vital after reaching the partial ionization zones (Figure\,\ref{fig-Ebind}). For the remnant mass between $0.764$ and $0.517\,M_\odot$, the binding energy $E_\mathrm{bind} = - \Delta E_1$ even becomes positive. This positive case is, at some level, similar to finding less massive TRGB donors \citep{2012ASPC..452....3H}. However, as the remnant mass approaches the donor's degenerate core, the binding energy difference between $E_\mathrm{bind}$ and $E^{'}_\mathrm{bind}$ becomes smaller again. Hence, the first two impacts discussed above only need to be concerned with donors with non-degenerate cores. For these short orbital sdB+WD binaries, we find the difference between CE efficiency parameters $\beta_\mathrm{CE}$ and $\alpha_\mathrm{CE}$ is relatively small (Figure\,\ref{fig-CEE-all}). Future binary population synthesis simulations of sdB binaries, applying a more accurate CE efficiency from fitted formulae \ref{beta-fit1} and \ref{beta-fit2} and previously found binding energy \citep{2001A&A...369..170T,2010ApJ...716..114X,2011ApJ...743...49L,2011MNRAS.411.2277D} are needed. 

\section{Comparison of CE Efficiency with Previous Studies}

The CEE outcome is essential to study the evolution and formation of short orbital-period compact binaries and double compact binaries. Energy or angular momentum conservation formulae can describe the CEE outcome. \citet{2008ASSL..352..233W} argues that any final energy state lower than the initial state requires the loss of angular momentum, and the converse is not necessarily true. So, \citet{2008ASSL..352..233W} concludes the energy budget more strongly constrains possible CEE outcomes.

The short orbital sdB+WD binaries offer the best examples to constrain CEE outcome because of the simplicity of sdB progenitors. Most sdB progenitors are located near the TRGB. Nevertheless, knowing the actual progenitor mass is necessary to constrain the CE efficiency parameter precisely. A short orbital-period sdB binary with a WD companion is likely formed after a first stable mass transfer phase and a second unstable CE phase. By applying a $P_\mathrm{orb}$--$M_\mathrm{WD}$ relation for the first stable mass transfer phase, we can precisely trace back the progenitor mass and evolutionary stage.

With an adequately (not just assuming $\lambda = 0.5$) calculated binding energy we determine the CE efficiency parameter of seven short orbital sdB+WD binaries with known inclination angles. Assuming a constant CE efficiency parameter $\beta_\mathrm{CE}$ exists its average is 0.32 (Figure~\ref{fig-CEE-all}). This average $\beta_\mathrm{CE}$ agrees with other findings from WD-binaries or WD+brwon-dwarf binaries \citep{2023MNRAS.518.3966S,2010A&A...520A..86Z,2022MNRAS.513.3587Z}. However, this CE efficiency might also be a statistical effect of the broad progenitor mass ranges in previous studies. We fit the CE efficiency parameter $\beta_\mathrm{CE}$ as a function of the initial and final binary masses (fitting formulae \ref{beta-fit1} and \ref{beta-fit2}).

\citet{2000A&A...360.1011N} find $\gamma$ between 1.4 and 1.7 for three helium WD systems based on their angular momentum description. Our findings are consistent with their results but with a slightly lower limit $1.78> \gamma > 1.20$ (Figure~\ref{fig-CEE-all}). \citet{2023ApJ...944...87D} study a large sample of post-CE WD+WD binaries based on their $\eta$ description. The $\eta$ in our study has the same trend as the results from \citet{2023ApJ...944...87D} but with a steeper gradient of 0.992 instead of 0.780. However, the difference is only noticeable for $\log_{10} (M_\mathrm{1e}/M_\mathrm{t}) < 0.1$, where almost has no WD+WD samples in the study by \citet{2023ApJ...944...87D}.

\section{Summary and Conclusion}

The CEE is an essential and long discussed physical stage in binary star evolution. Its outcome is related to the evolution and formation of many short orbital-period compact binaries, including the progenitors of merging stellar gravitational wave sources. Short orbital-period sdB+WD binaries, formed after a first stable mass transfer phase and a second unstable CE phase, are excellent samples to study the fundamental process of the CEE. By applying a $P_\mathrm{orb}$--$M_\mathrm{WD}$ relation from the first stable mass transfer process, the progenitor mass and evolutionary stage of the current sdB can be precisely constrained. A sound $P_\mathrm{orb}$--$M_\mathrm{WD}$ relation is, in priciple, suitable to constrain the CEE outcome of other WD binaries.

By assuming an adiabatic CE ejection process, we trace the donor's total energy change as a function of its remnant mass. The response of the stellar interior to mass loss is naturally addressed. We use a total energy conservation assumption, plus the need for both components to shrink back within their Roche lobes, to constrain the CEE outcome of seven short orbital period sdB+WD binaries. The average CE efficiency parameters in this study are consistent with previous studies \citep{2010A&A...520A..86Z,2022MNRAS.517.2867H,2022MNRAS.513.3587Z,2023MNRAS.518.3966S}. With well-constrained sdB progenitor masses and evolutionary stages, the CE efficiency parameter is fitted as a function of the initial mass ratio (Eqation~\ref{beta-fit1}) and other forms (Equation~\ref{beta-fit2}). In addition to energy descriptions, we also discuss the results from angular momentum descriptions.

The impact of different binding energy calculations on the CE efficiency parameter is small. This result is as we expected because the envelope mass of an sdB is tiny. However, the impact would become more critical for other post-CE binaries with more massive progenitors, especially those with none-degenerate cores. The binding energy is also sensitive to the internal energy. For relatively more massive progenitors in these seven sdB+WD binaries the thermal efficiency $\alpha_\mathrm{th}$ needs to be close to 1.

Acknowledgments

We wish to thank the referee for his valuable comments and suggestions, which have helped us further improve this work. This project is supported by the National Natural Science Foundation of China (NSFC, grant Nos. 12288102, 12125303), National Key R\&D Program of China (2021YFA1600403, 2021YFA1600401), NSFC (grant Nos. 12090040/3, 12173081), Yunnan Fundamental Research Projects (grant NOs. 202101AV070001), Yunnan Revitalization Talent Support Program - Science \& Technology Champion Project (No. 202305AB350003), and International Centre of Supernovae, Yunnan Key Laboratory (No. 202302AN360001), the Key Research Program of Frontier Sciences, CAS, No. ZDBS-LY-7005, CAS, Light of West China Program and International Centre of Supernovae, Yunnan Key Laboratory (No. 202302AN360001). CAT thanks Churchill College for his fellowship. AS thanks the Gates Cambridge Trust for his scholarship.



\bibliography{hongwei8}{}

\begin{thebibliography}{}
\expandafter\ifx\csname natexlab\endcsname\relax\def\natexlab#1{#1}\fi
\providecommand{\url}[1]{\href{#1}{#1}}
\providecommand{\dodoi}[1]{doi:~\href{http://doi.org/#1}{\nolinkurl{#1}}}
\providecommand{\doeprint}[1]{\href{http://ascl.net/#1}{\nolinkurl{http://ascl.net/#1}}}
\providecommand{\doarXiv}[1]{\href{https://arxiv.org/abs/#1}{\nolinkurl{https://arxiv.org/abs/#1}}}

\bibitem[{{Belczynski} {et~al.}(2002){Belczynski}, {Kalogera}, \&
  {Bulik}}]{2002ApJ...572..407B}
{Belczynski}, K., {Kalogera}, V., \& {Bulik}, T. 2002, \apj, 572, 407,
  \dodoi{10.1086/340304}

\bibitem[{{Bhattacharya} \& {van den Heuvel}(1991)}]{1991PhR...203....1B}
{Bhattacharya}, D., \& {van den Heuvel}, E.~P.~J. 1991, \physrep, 203, 1,
  \dodoi{10.1016/0370-1573(91)90064-S}

\bibitem[{{Broekgaarden} {et~al.}(2021){Broekgaarden}, {Berger}, {Neijssel},
  {Vigna-G{\'o}mez}, {Chattopadhyay}, {Stevenson}, {Chruslinska}, {Justham},
  {de Mink}, \& {Mandel}}]{2021MNRAS.508.5028B}
{Broekgaarden}, F.~S., {Berger}, E., {Neijssel}, C.~J., {et~al.} 2021, \mnras,
  508, 5028, \dodoi{10.1093/mnras/stab2716}

\bibitem[{{Cai} {et~al.}(2022){Cai}, {Pastorello}, {Fraser}, {Wang},
  {Filippenko}, {Reguitti}, {Patra}, {Goranskij}, {Barsukova}, {Brink},
  {Elias-Rosa}, {Stevance}, {Zheng}, {Yang}, {Atapin}, {Benetti}, {de Boer},
  {Bose}, {Burke}, {Byrne}, {Cappellaro}, {Chambers}, {Chen}, {Emami}, {Gao},
  {Hiramatsu}, {Howell}, {Huber}, {Kankare}, {Kelly}, {Kotak}, {Kravtsov},
  {Lander}, {Li}, {Lin}, {Lundqvist}, {Magnier}, {Malygin}, {Maslennikova},
  {Matilainen}, {Mazzali}, {McCully}, {Mo}, {Moran}, {Newsome}, {Oparin},
  {Padilla Gonzalez}, {Reynolds}, {Shatsky}, {Smartt}, {Smith}, {Stritzinger},
  {Tatarnikov}, {Terreran}, {Uklein}, {Valerin}, {Vallely}, {Vozyakova},
  {Wainscoat}, {Yan}, {Zhang}, {Zhang}, {Zheltoukhov}, {Dastidar}, {Fulton},
  {Galbany}, {Gangopadhyay}, {Ge}, {Guti{\'e}rrez}, {Lin}, {Misra}, {Ou},
  {Salmaso}, {Tartaglia}, {Xiao}, \& {Zhang}}]{2022A&A...667A...4C}
{Cai}, Y.~Z., {Pastorello}, A., {Fraser}, M., {et~al.} 2022, \aap, 667, A4,
  \dodoi{10.1051/0004-6361/202244393}

\bibitem[{{Chen} \& {Liu}(2013)}]{2013MNRAS.432L..75C}
{Chen}, W.~C., \& {Liu}, W.~M. 2013, \mnras, 432, L75,
  \dodoi{10.1093/mnrasl/slt043}

\bibitem[{{Chen} {et~al.}(2013){Chen}, {Han}, {Deca}, \&
  {Podsiadlowski}}]{2013MNRAS.434..186C}
{Chen}, X., {Han}, Z., {Deca}, J., \& {Podsiadlowski}, P. 2013, \mnras, 434,
  186, \dodoi{10.1093/mnras/stt992}

\bibitem[{{Copperwheat} {et~al.}(2011){Copperwheat}, {Morales-Rueda}, {Marsh},
  {Maxted}, \& {Heber}}]{2011MNRAS.415.1381C}
{Copperwheat}, C.~M., {Morales-Rueda}, L., {Marsh}, T.~R., {Maxted}, P.~F.~L.,
  \& {Heber}, U. 2011, \mnras, 415, 1381,
  \dodoi{10.1111/j.1365-2966.2011.18786.x}

\bibitem[{{de Kool}(1990)}]{1990ApJ...358..189D}
{de Kool}, M. 1990, \apj, 358, 189, \dodoi{10.1086/168974}

\bibitem[{{De Marco} {et~al.}(2011){De Marco}, {Passy}, {Moe}, {Herwig}, {Mac
  Low}, \& {Paxton}}]{2011MNRAS.411.2277D}
{De Marco}, O., {Passy}, J.-C., {Moe}, M., {et~al.} 2011, \mnras, 411, 2277,
  \dodoi{10.1111/j.1365-2966.2010.17891.x}

\bibitem[{{Di Stefano} {et~al.}(2023){Di Stefano}, {Kruckow}, {Gao},
  {Neunteufel}, \& {Kobayashi}}]{2023ApJ...944...87D}
{Di Stefano}, R., {Kruckow}, M.~U., {Gao}, Y., {Neunteufel}, P.~G., \&
  {Kobayashi}, C. 2023, \apj, 944, 87, \dodoi{10.3847/1538-4357/acae9b}

\bibitem[{{Eggleton}(1983)}]{1983ApJ...268..368E}
{Eggleton}, P.~P. 1983, \apj, 268, 368, \dodoi{10.1086/160960}

\bibitem[{{Gallegos-Garcia} {et~al.}(2023){Gallegos-Garcia}, {Berry}, \&
  {Kalogera}}]{2023ApJ...955..133G}
{Gallegos-Garcia}, M., {Berry}, C. P.~L., \& {Kalogera}, V. 2023, \apj, 955,
  133, \dodoi{10.3847/1538-4357/ace434}

\bibitem[{{Gao} \& {Li}(2023)}]{2023MNRAS.525.2605G}
{Gao}, S.-J., \& {Li}, X.-D. 2023, \mnras, 525, 2605,
  \dodoi{10.1093/mnras/stad2446}

\bibitem[{{Ge} {et~al.}(2010{\natexlab{a}}){Ge}, {Hjellming}, {Webbink},
  {Chen}, \& {Han}}]{2010ApJ...717..724G}
{Ge}, H., {Hjellming}, M.~S., {Webbink}, R.~F., {Chen}, X., \& {Han}, Z.
  2010{\natexlab{a}}, \apj, 717, 724, \dodoi{10.1088/0004-637X/717/2/724}

\bibitem[{{Ge} {et~al.}(2023){Ge}, {Tout}, {Chen}, {Sarkar}, {Walton}, \&
  {Han}}]{2023ApJ...945....7G}
{Ge}, H., {Tout}, C.~A., {Chen}, X., {et~al.} 2023, \apj, 945, 7,
  \dodoi{10.3847/1538-4357/acb7e9}

\bibitem[{{Ge} {et~al.}(2015){Ge}, {Webbink}, {Chen}, \&
  {Han}}]{2015ApJ...812...40G}
{Ge}, H., {Webbink}, R.~F., {Chen}, X., \& {Han}, Z. 2015, \apj, 812, 40,
  \dodoi{10.1088/0004-637X/812/1/40}

\bibitem[{{Ge} {et~al.}(2020{\natexlab{a}}){Ge}, {Webbink}, {Chen}, \&
  {Han}}]{2020ApJ...899..132G}
---. 2020{\natexlab{a}}, \apj, 899, 132, \dodoi{10.3847/1538-4357/aba7b7}

\bibitem[{{Ge} {et~al.}(2020{\natexlab{b}}){Ge}, {Webbink}, \&
  {Han}}]{2020ApJS..249....9G}
{Ge}, H., {Webbink}, R.~F., \& {Han}, Z. 2020{\natexlab{b}}, \apjs, 249, 9,
  \dodoi{10.3847/1538-4365/ab98f6}

\bibitem[{{Ge} {et~al.}(2010{\natexlab{b}}){Ge}, {Webbink}, {Han}, \&
  {Chen}}]{2010ApSS.329..243G}
{Ge}, H., {Webbink}, R.~F., {Han}, Z., \& {Chen}, X. 2010{\natexlab{b}}, \apss,
  329, 243, \dodoi{10.1007/s10509-010-0286-1}

\bibitem[{{Ge} {et~al.}(2022){Ge}, {Tout}, {Chen}, {Kruckow}, {Chen}, {Jiang},
  {Li}, {Liu}, \& {Han}}]{2022ApJ...933..137G}
{Ge}, H., {Tout}, C.~A., {Chen}, X., {et~al.} 2022, \apj, 933, 137,
  \dodoi{10.3847/1538-4357/ac75d3}

\bibitem[{{Geier}(2020)}]{2020A&A...635A.193G}
{Geier}, S. 2020, \aap, 635, A193, \dodoi{10.1051/0004-6361/202037526}

\bibitem[{{Geier} {et~al.}(2010){Geier}, {Heber}, {Kupfer}, \&
  {Napiwotzki}}]{2010A-A...515A..37G}
{Geier}, S., {Heber}, U., {Kupfer}, T., \& {Napiwotzki}, R. 2010, \aap, 515,
  A37, \dodoi{10.1051/0004-6361/200912545}

\bibitem[{{Geier} {et~al.}(2011){Geier}, {Hirsch}, {Tillich}, {Maxted},
  {Bentley}, {{\O}stensen}, {Heber}, {G{\"a}nsicke}, {Marsh}, {Napiwotzki},
  {Barlow}, \& {O'Toole}}]{2011A&A...530A..28G}
{Geier}, S., {Hirsch}, H., {Tillich}, A., {et~al.} 2011, \aap, 530, A28,
  \dodoi{10.1051/0004-6361/201015316}

\bibitem[{{Grichener}(2023)}]{2023MNRAS.523..221G}
{Grichener}, A. 2023, \mnras, 523, 221, \dodoi{10.1093/mnras/stad1449}

\bibitem[{{Han} {et~al.}(2012){Han}, {Chen}, {Lei}, \&
  {Podsiadlowski}}]{2012ASPC..452....3H}
{Han}, Z., {Chen}, X., {Lei}, Z., \& {Podsiadlowski}, P. 2012, in Astronomical
  Society of the Pacific Conference Series, Vol. 452, Fifth Meeting on Hot
  Subdwarf Stars and Related Objects, ed. D.~{Kilkenny}, C.~S. {Jeffery}, \&
  C.~{Koen}, 3

\bibitem[{{Han} {et~al.}(1995){Han}, {Podsiadlowski}, \&
  {Eggleton}}]{1995MNRAS.272..800H}
{Han}, Z., {Podsiadlowski}, P., \& {Eggleton}, P.~P. 1995, \mnras, 272, 800,
  \dodoi{10.1093/mnras/272.4.800}

\bibitem[{{Han} {et~al.}(2003){Han}, {Podsiadlowski}, {Maxted}, \&
  {Marsh}}]{2003MNRAS.341..669H}
{Han}, Z., {Podsiadlowski}, P., {Maxted}, P.~F.~L., \& {Marsh}, T.~R. 2003,
  \mnras, 341, 669, \dodoi{10.1046/j.1365-8711.2003.06451.x}

\bibitem[{{Han} {et~al.}(2002){Han}, {Podsiadlowski}, {Maxted}, {Marsh}, \&
  {Ivanova}}]{2002MNRAS.336..449H}
{Han}, Z., {Podsiadlowski}, P., {Maxted}, P.~F.~L., {Marsh}, T.~R., \&
  {Ivanova}, N. 2002, \mnras, 336, 449,
  \dodoi{10.1046/j.1365-8711.2002.05752.x}

\bibitem[{{Heber}(2009)}]{2009ARA&A..47..211H}
{Heber}, U. 2009, \araa, 47, 211, \dodoi{10.1146/annurev-astro-082708-101836}

\bibitem[{{Heber}(2016)}]{2016PASP..128h2001H}
---. 2016, \pasp, 128, 082001, \dodoi{10.1088/1538-3873/128/966/082001}

\bibitem[{{Hernandez} {et~al.}(2022){Hernandez}, {Schreiber}, {Parsons},
  {G{\"a}nsicke}, {Toloza}, {Zorotovic}, {Raddi}, {Rebassa-Mansergas}, \&
  {Ren}}]{2022MNRAS.517.2867H}
{Hernandez}, M.~S., {Schreiber}, M.~R., {Parsons}, S.~G., {et~al.} 2022,
  \mnras, 517, 2867, \dodoi{10.1093/mnras/stac2837}

\bibitem[{{Iben} \& {Tutukov}(1984)}]{1984ApJS...54..335I}
{Iben}, I., J., \& {Tutukov}, A.~V. 1984, \apjs, 54, 335,
  \dodoi{10.1086/190932}

\bibitem[{{Iorio} {et~al.}(2023){Iorio}, {Mapelli}, {Costa}, {Spera},
  {Escobar}, {Sgalletta}, {Trani}, {Korb}, {Santoliquido}, {Dall'Amico},
  {Gaspari}, \& {Bressan}}]{2023MNRAS.524..426I}
{Iorio}, G., {Mapelli}, M., {Costa}, G., {et~al.} 2023, \mnras, 524, 426,
  \dodoi{10.1093/mnras/stad1630}

\bibitem[{{Ivanova} {et~al.}(2001){Ivanova}, {Podsiadlowski}, \&
  {Spruit}}]{2001ASPC..229..261I}
{Ivanova}, N., {Podsiadlowski}, P., \& {Spruit}, H. 2001, in Astronomical
  Society of the Pacific Conference Series, Vol. 229, Evolution of Binary and
  Multiple Star Systems, ed. P.~{Podsiadlowski}, S.~{Rappaport}, A.~R. {King},
  F.~{D'Antona}, \& L.~{Burderi}, 261, \dodoi{10.48550/arXiv.astro-ph/0102141}

\bibitem[{{Ivanova} {et~al.}(2013){Ivanova}, {Justham}, {Chen}, {De Marco},
  {Fryer}, {Gaburov}, {Ge}, {Glebbeek}, {Han}, {Li}, {Lu}, {Marsh},
  {Podsiadlowski}, {Potter}, {Soker}, {Taam}, {Tauris}, {van den Heuvel}, \&
  {Webbink}}]{2013A&ARv..21...59I}
{Ivanova}, N., {Justham}, S., {Chen}, X., {et~al.} 2013, \aapr, 21, 59,
  \dodoi{10.1007/s00159-013-0059-2}

\bibitem[{{Joss} {et~al.}(1987){Joss}, {Rappaport}, \&
  {Lewis}}]{1987ApJ...319..180J}
{Joss}, P.~C., {Rappaport}, S., \& {Lewis}, W. 1987, \apj, 319, 180,
  \dodoi{10.1086/165443}

\bibitem[{{Kashi} \& {Soker}(2011)}]{2011MNRAS.417.1466K}
{Kashi}, A., \& {Soker}, N. 2011, \mnras, 417, 1466,
  \dodoi{10.1111/j.1365-2966.2011.19361.x}

\bibitem[{{Kawka} {et~al.}(2015){Kawka}, {Vennes}, {O'Toole}, {N{\'e}meth},
  {Burton}, {Kotze}, \& {Buckley}}]{2015MNRAS.450.3514K}
{Kawka}, A., {Vennes}, S., {O'Toole}, S., {et~al.} 2015, \mnras, 450, 3514,
  \dodoi{10.1093/mnras/stv821}

\bibitem[{{Kippenhahn} \& {Weigert}(1967)}]{1967ZA.....65..251K}
{Kippenhahn}, R., \& {Weigert}, A. 1967, \zap, 65, 251

\bibitem[{{Kupfer} {et~al.}(2015){Kupfer}, {Geier}, {Heber}, {{\O}stensen},
  {Barlow}, {Maxted}, {Heuser}, {Schaffenroth}, \&
  {G{\"a}nsicke}}]{2015A&A...576A..44K}
{Kupfer}, T., {Geier}, S., {Heber}, U., {et~al.} 2015, \aap, 576, A44,
  \dodoi{10.1051/0004-6361/201425213}

\bibitem[{{Lei} {et~al.}(2023){Lei}, {He}, {N{\'e}meth}, {Vos}, {Zou}, {Hu},
  {Xiao}, {Yan}, \& {Zhao}}]{2023ApJ...942..109L}
{Lei}, Z., {He}, R., {N{\'e}meth}, P., {et~al.} 2023, \apj, 942, 109,
  \dodoi{10.3847/1538-4357/aca542}

\bibitem[{{Li} {et~al.}(2023){Li}, {Chen}, {Ge}, {Chen}, \&
  {Han}}]{2023A&A...669A..82L}
{Li}, Z., {Chen}, X., {Ge}, H., {Chen}, H.-L., \& {Han}, Z. 2023, \aap, 669,
  A82, \dodoi{10.1051/0004-6361/202243893}

\bibitem[{{Liu} {et~al.}(2018){Liu}, {Wang}, {Chen}, {Zuo}, \&
  {Han}}]{2018MNRAS.477..384L}
{Liu}, D., {Wang}, B., {Chen}, W., {Zuo}, Z., \& {Han}, Z. 2018, \mnras, 477,
  384, \dodoi{10.1093/mnras/sty561}

\bibitem[{{Livio} \& {Soker}(1988)}]{1988ApJ...329..764L}
{Livio}, M., \& {Soker}, N. 1988, \apj, 329, 764, \dodoi{10.1086/166419}

\bibitem[{{Loveridge} {et~al.}(2011){Loveridge}, {van der Sluys}, \&
  {Kalogera}}]{2011ApJ...743...49L}
{Loveridge}, A.~J., {van der Sluys}, M.~V., \& {Kalogera}, V. 2011, \apj, 743,
  49, \dodoi{10.1088/0004-637X/743/1/49}

\bibitem[{{Luo} {et~al.}(2021){Luo}, {N{\'e}meth}, {Wang}, {Wang}, \&
  {Han}}]{2021ApJS..256...28L}
{Luo}, Y., {N{\'e}meth}, P., {Wang}, K., {Wang}, X., \& {Han}, Z. 2021, \apjs,
  256, 28, \dodoi{10.3847/1538-4365/ac11f6}

\bibitem[{{Matsumoto} \& {Metzger}(2022)}]{2022ApJ...938....5M}
{Matsumoto}, T., \& {Metzger}, B.~D. 2022, \apj, 938, 5,
  \dodoi{10.3847/1538-4357/ac6269}

\bibitem[{{Maxted} {et~al.}(2001){Maxted}, {Heber}, {Marsh}, \&
  {North}}]{2001MNRAS.326.1391M}
{Maxted}, P.~F.~L., {Heber}, U., {Marsh}, T.~R., \& {North}, R.~C. 2001,
  \mnras, 326, 1391, \dodoi{10.1111/j.1365-2966.2001.04714.x}

\bibitem[{{Metzger}(2022)}]{2022ApJ...932...84M}
{Metzger}, B.~D. 2022, \apj, 932, 84, \dodoi{10.3847/1538-4357/ac6d59}

\bibitem[{{Nelemans} \& {Tout}(2005)}]{2005MNRAS.356..753N}
{Nelemans}, G., \& {Tout}, C.~A. 2005, \mnras, 356, 753,
  \dodoi{10.1111/j.1365-2966.2004.08496.x}

\bibitem[{{Nelemans} {et~al.}(2000){Nelemans}, {Verbunt}, {Yungelson}, \&
  {Portegies Zwart}}]{2000A&A...360.1011N}
{Nelemans}, G., {Verbunt}, F., {Yungelson}, L.~R., \& {Portegies Zwart}, S.~F.
  2000, \aap, 360, 1011, \dodoi{10.48550/arXiv.astro-ph/0006216}

\bibitem[{{Olejak} {et~al.}(2021){Olejak}, {Belczynski}, \&
  {Ivanova}}]{2021A&A...651A.100O}
{Olejak}, A., {Belczynski}, K., \& {Ivanova}, N. 2021, \aap, 651, A100,
  \dodoi{10.1051/0004-6361/202140520}

\bibitem[{{Paczy{\'n}ski}(1971{\natexlab{a}})}]{1971AcA....21..417P}
{Paczy{\'n}ski}, B. 1971{\natexlab{a}}, \actaa, 21, 417

\bibitem[{{Paczy{\'n}ski}(1971{\natexlab{b}})}]{1971ARA&A...9..183P}
---. 1971{\natexlab{b}}, \araa, 9, 183,
  \dodoi{10.1146/annurev.aa.09.090171.001151}

\bibitem[{{Paczynski}(1976)}]{1976IAUS...73...75P}
{Paczynski}, B. 1976, in Structure and Evolution of Close Binary Systems, ed.
  P.~{Eggleton}, S.~{Mitton}, \& J.~{Whelan}, Vol.~73, 75

\bibitem[{{Podsiadlowski} {et~al.}(2008){Podsiadlowski}, {Han}, {Lynas-Gray},
  \& {Brown}}]{2008ASPC..392...15P}
{Podsiadlowski}, P., {Han}, Z., {Lynas-Gray}, A.~E., \& {Brown}, D. 2008, in
  Astronomical Society of the Pacific Conference Series, Vol. 392, Hot Subdwarf
  Stars and Related Objects, ed. U.~{Heber}, C.~S. {Jeffery}, \&
  R.~{Napiwotzki}, 15, \dodoi{10.48550/arXiv.0808.0574}

\bibitem[{{Podsiadlowski} \& {Rappaport}(2000)}]{2000ApJ...529..946P}
{Podsiadlowski}, P., \& {Rappaport}, S. 2000, \apj, 529, 946,
  \dodoi{10.1086/308323}

\bibitem[{{Politano} {et~al.}(2008){Politano}, {Taam}, {van der Sluys}, \&
  {Willems}}]{2008ApJ...687L..99P}
{Politano}, M., {Taam}, R.~E., {van der Sluys}, M., \& {Willems}, B. 2008,
  \apjl, 687, L99, \dodoi{10.1086/593328}

\bibitem[{{Pols} {et~al.}(1995){Pols}, {Tout}, {Eggleton}, \&
  {Han}}]{1995MNRAS.274..964P}
{Pols}, O.~R., {Tout}, C.~A., {Eggleton}, P.~P., \& {Han}, Z. 1995, \mnras,
  274, 964, \dodoi{10.1093/mnras/274.3.964}

\bibitem[{{Rappaport} {et~al.}(1995){Rappaport}, {Podsiadlowski}, {Joss}, {Di
  Stefano}, \& {Han}}]{1995MNRAS.273..731R}
{Rappaport}, S., {Podsiadlowski}, P., {Joss}, P.~C., {Di Stefano}, R., \&
  {Han}, Z. 1995, \mnras, 273, 731, \dodoi{10.1093/mnras/273.3.731}

\bibitem[{{Refsdal} \& {Weigert}(1971)}]{1971A&A....13..367R}
{Refsdal}, S., \& {Weigert}, A. 1971, \aap, 13, 367

\bibitem[{{R{\"o}pke} \& {De Marco}(2023)}]{2023LRCA....9....2R}
{R{\"o}pke}, F.~K., \& {De Marco}, O. 2023, Living Reviews in Computational
  Astrophysics, 9, 2, \dodoi{10.1007/s41115-023-00017-x}

\bibitem[{{Saio} \& {Jeffery}(2000)}]{2000MNRAS.313..671S}
{Saio}, H., \& {Jeffery}, C.~S. 2000, \mnras, 313, 671,
  \dodoi{10.1046/j.1365-8711.2000.03221.x}

\bibitem[{{Schaffenroth} {et~al.}(2023){Schaffenroth}, {Barlow}, {Pelisoli},
  {Geier}, \& {Kupfer}}]{2023A-A...673A..90S}
{Schaffenroth}, V., {Barlow}, B.~N., {Pelisoli}, I., {Geier}, S., \& {Kupfer},
  T. 2023, \aap, 673, A90, \dodoi{10.1051/0004-6361/202244697}

\bibitem[{{Schaffenroth} {et~al.}(2022){Schaffenroth}, {Pelisoli}, {Barlow},
  {Geier}, \& {Kupfer}}]{2022A&A...666A.182S}
{Schaffenroth}, V., {Pelisoli}, I., {Barlow}, B.~N., {Geier}, S., \& {Kupfer},
  T. 2022, \aap, 666, A182, \dodoi{10.1051/0004-6361/202244214}

\bibitem[{{Scherbak} \& {Fuller}(2023)}]{2023MNRAS.518.3966S}
{Scherbak}, P., \& {Fuller}, J. 2023, \mnras, 518, 3966,
  \dodoi{10.1093/mnras/stac3313}

\bibitem[{{Shao} \& {Li}(2021)}]{2021ApJ...920...81S}
{Shao}, Y., \& {Li}, X.-D. 2021, \apj, 920, 81,
  \dodoi{10.3847/1538-4357/ac173e}

\bibitem[{{Smedley} {et~al.}(2014){Smedley}, {Tout}, {Ferrario}, \&
  {Wickramasinghe}}]{2014MNRAS.437.2217S}
{Smedley}, S.~L., {Tout}, C.~A., {Ferrario}, L., \& {Wickramasinghe}, D.~T.
  2014, \mnras, 437, 2217, \dodoi{10.1093/mnras/stt2030}

\bibitem[{{Soker}(1998)}]{1998AJ....116.1308S}
{Soker}, N. 1998, \aj, 116, 1308, \dodoi{10.1086/300503}

\bibitem[{{Soker}(2017)}]{2017MNRAS.470L.102S}
---. 2017, \mnras, 470, L102, \dodoi{10.1093/mnrasl/slx089}

\bibitem[{{Taam} \& {Sandquist}(2000)}]{2000ARA&A..38..113T}
{Taam}, R.~E., \& {Sandquist}, E.~L. 2000, \araa, 38, 113,
  \dodoi{10.1146/annurev.astro.38.1.113}

\bibitem[{{Tauris} \& {Dewi}(2001)}]{2001A&A...369..170T}
{Tauris}, T.~M., \& {Dewi}, J.~D.~M. 2001, \aap, 369, 170,
  \dodoi{10.1051/0004-6361:20010099}

\bibitem[{{Tauris} \& {Savonije}(1999)}]{1999A&A...350..928T}
{Tauris}, T.~M., \& {Savonije}, G.~J. 1999, \aap, 350, 928,
  \dodoi{10.48550/arXiv.astro-ph/9909147}

\bibitem[{{Tauris} \& {van den Heuvel}(2023)}]{2023pbse.book.....T}
{Tauris}, T.~M., \& {van den Heuvel}, E. P.~J. 2023, {Physics of Binary Star
  Evolution. From Stars to X-ray Binaries and Gravitational Wave Sources},
  \dodoi{10.48550/arXiv.2305.09388}

\bibitem[{{Tauris} {et~al.}(2000){Tauris}, {van den Heuvel}, \&
  {Savonije}}]{2000ApJ...530L..93T}
{Tauris}, T.~M., {van den Heuvel}, E. P.~J., \& {Savonije}, G.~J. 2000, \apjl,
  530, L93, \dodoi{10.1086/312496}

\bibitem[{{Toonen} {et~al.}(2016){Toonen}, {Hamers}, \& {Portegies
  Zwart}}]{2016ComAC...3....6T}
{Toonen}, S., {Hamers}, A., \& {Portegies Zwart}, S. 2016, Computational
  Astrophysics and Cosmology, 3, 6, \dodoi{10.1186/s40668-016-0019-0}

\bibitem[{{Tout} \& {Eggleton}(1988)}]{1988MNRAS.231..823T}
{Tout}, C.~A., \& {Eggleton}, P.~P. 1988, \mnras, 231, 823,
  \dodoi{10.1093/mnras/231.4.823}

\bibitem[{{Vos} {et~al.}(2018){Vos}, {N{\'e}meth}, {Vu{\v{c}}kovi{\'c}},
  {{\O}stensen}, \& {Parsons}}]{2018MNRAS.473..693V}
{Vos}, J., {N{\'e}meth}, P., {Vu{\v{c}}kovi{\'c}}, M., {{\O}stensen}, R., \&
  {Parsons}, S. 2018, \mnras, 473, 693, \dodoi{10.1093/mnras/stx2198}

\bibitem[{{Wang}(2018)}]{2018RAA....18...49W}
{Wang}, B. 2018, Research in Astronomy and Astrophysics, 18, 049,
  \dodoi{10.1088/1674-4527/18/5/49}

\bibitem[{{Webbink}(1984)}]{1984ApJ...277..355W}
{Webbink}, R.~F. 1984, \apj, 277, 355, \dodoi{10.1086/161701}

\bibitem[{{Webbink}(2008)}]{2008ASSL..352..233W}
{Webbink}, R.~F. 2008, in Astrophysics and Space Science Library, Vol. 352,
  Astrophysics and Space Science Library, ed. E.~F. {Milone}, D.~A. {Leahy}, \&
  D.~W. {Hobill}, 233, \dodoi{10.1007/978-1-4020-6544-6_13}

\bibitem[{{Webbink} {et~al.}(1983){Webbink}, {Rappaport}, \&
  {Savonije}}]{1983ApJ...270..678W}
{Webbink}, R.~F., {Rappaport}, S., \& {Savonije}, G.~J. 1983, \apj, 270, 678,
  \dodoi{10.1086/161159}

\bibitem[{{Xu} \& {Li}(2010)}]{2010ApJ...716..114X}
{Xu}, X.-J., \& {Li}, X.-D. 2010, \apj, 716, 114,
  \dodoi{10.1088/0004-637X/716/1/114}

\bibitem[{{Zhang} \& {Jeffery}(2012)}]{2012MNRAS.419..452Z}
{Zhang}, X., \& {Jeffery}, C.~S. 2012, \mnras, 419, 452,
  \dodoi{10.1111/j.1365-2966.2011.19711.x}

\bibitem[{{Zhang} {et~al.}(2021{\natexlab{a}}){Zhang}, {Chen}, {Chen}, \&
  {Han}}]{2021MNRAS.502..383Z}
{Zhang}, Y., {Chen}, H., {Chen}, X., \& {Han}, Z. 2021{\natexlab{a}}, \mnras,
  502, 383, \dodoi{10.1093/mnras/stab020}

\bibitem[{{Zhang} {et~al.}(2021{\natexlab{b}}){Zhang}, {Chen}, {Xiong}, {Chen},
  \& {Han}}]{2021MNRAS.505.3514Z}
{Zhang}, Y., {Chen}, H.-L., {Xiong}, H., {Chen}, X., \& {Han}, Z.
  2021{\natexlab{b}}, \mnras, 505, 3514, \dodoi{10.1093/mnras/stab1627}

\bibitem[{{Zorotovic} \& {Schreiber}(2022)}]{2022MNRAS.513.3587Z}
{Zorotovic}, M., \& {Schreiber}, M. 2022, \mnras, 513, 3587,
  \dodoi{10.1093/mnras/stac1137}

\bibitem[{{Zorotovic} {et~al.}(2010){Zorotovic}, {Schreiber}, {G{\"a}nsicke},
  \& {Nebot G{\'o}mez-Mor{\'a}n}}]{2010A&A...520A..86Z}
{Zorotovic}, M., {Schreiber}, M.~R., {G{\"a}nsicke}, B.~T., \& {Nebot
  G{\'o}mez-Mor{\'a}n}, A. 2010, \aap, 520, A86,
  \dodoi{10.1051/0004-6361/200913658}

\end{thebibliography}
\bibliographystyle{aasjournal}

\listofchanges

\end{document}